\documentclass{article}

\usepackage{microtype}
\usepackage{graphicx}
\usepackage{subfigure}
\usepackage{booktabs} 

\usepackage{hyperref}

\usepackage[accepted]{mlsys2022}

\mlsystitlerunning{dPRO: A Generic Profiling and Optimization System for Expediting Distributed DNN Training}

\usepackage{tikz}
\usepackage{amsmath}
\usepackage{algorithm}
\usepackage{color,soul}
\usepackage{subfigure}
\usepackage[normalem]{ulem}
\usepackage{multirow}
\usepackage{makecell}
\usepackage{comment}
\usepackage{hyperref}

\usepackage{listings}
\definecolor{dkgreen}{rgb}{0,0.6,0}
\definecolor{gray}{rgb}{0.5,0.5,0.5}
\definecolor{mauve}{rgb}{0.58,0,0.82}
\usepackage{pifont}

\def\eg{\textit{e.g.}}
\def\ie{\textit{i.e.}}
\newcommand{\mypara}[1]{\noindent{\bf {#1}}~}
\newcommand{\Hu}[1]{\textcolor{cyan}{[Hanpeng: #1]}}
\newcommand{\CY}[1]{\textcolor{blue}{[CY: #1]}}
\newcommand{\yanghua}[1]{\textcolor{brown}{[Yanghua: #1]}}

\newcommand{\cwu}[1]{\textcolor{purple}{[CWu: #1]}}
\newcommand{\ignore}[1]{}

\newcommand{\sysname}{dPRO}
\newcommand{\mutator}{Graph Pass}
\newcommand{\mutators}{Graph Passes}
\newcommand{\layerview}{Coarsened View}
\newcommand{\layername}{group}

\newtheorem{theorem}{Theorem} 

\begin{document}

\twocolumn[
\mlsystitle{\Large \bf dPRO: A Generic Performance Diagnosis and Optimization Toolkit for Expediting Distributed DNN Training}



\mlsyssetsymbol{equal}{*}

\begin{mlsysauthorlist}
\mlsysauthor{Hanpeng Hu}{hk,bd}
\mlsysauthor{Chenyu Jiang}{hk}
\mlsysauthor{Yuchen Zhong}{hk}
\mlsysauthor{Yanghua Peng}{bd}
\mlsysauthor{Chuan Wu}{hk}
\mlsysauthor{Yibo Zhu}{bd}
\mlsysauthor{Haibin Lin}{bd}
\mlsysauthor{Chuanxiong Guo}{bd}
\end{mlsysauthorlist}

\mlsysaffiliation{hk}{Department of Computer Science, University of Hong Kong, Hong Kong, China}
\mlsysaffiliation{bd}{ByteDance Inc., Beijing, China}

\mlsyscorrespondingauthor{Hanpeng Hu}{hphu@cs.hku.hk}

\mlsyskeywords{Machine Learning, MLSys, Distributed DNN Training, ML Optimization}

\vskip 0.3in

\begin{abstract}

Distributed training using multiple devices (\eg, GPUs) has been widely adopted for learning DNN models over large datasets. However, the performance of large-scale distributed training tends to be far from linear speed-up in practice. Given the complexity of distributed systems, it is challenging to identify the root cause(s) of inefficiency and exercise effective performance optimizations when unexpected low training speed occurs. 
To date, there exists no software tool which diagnoses performance issues and helps expedite distributed DNN training, while the training can be run using different deep learning frameworks. This paper proposes {\sysname}, a toolkit that includes: (1) an efficient profiler that collects runtime traces of distributed DNN training across multiple frameworks, especially fine-grained communication traces, and constructs global data flow graphs including detailed communication operations for accurate replay; 
(2) an optimizer that effectively identifies performance bottlenecks and explores optimization 
strategies (from computation, communication, and memory aspects) for training acceleration. We implement {\sysname} on multiple deep learning frameworks (TensorFlow, MXNet) and representative communication schemes (AllReduce and Parameter Server). Extensive experiments show that  {\sysname} predicts the performance of distributed training in various settings with {$<5\%$} errors in most cases and finds optimization strategies with up to {$3.48 \times$} speed-up over the baselines. 
\end{abstract}
]



\printAffiliationsAndNotice{\mlsysEqualContribution} 

\section{Introduction}
Distributed training on large datasets has been widely adopted to learn modern machine learning (ML) models such as deep neural networks (DNNs), to power various AI-driven applications. 
As compared to single-node training, distributed training using devices on multiple servers substantially alleviates the time cost \cite{huang2019gpipe,
lepikhin2021gshard}, 
but often fails to 
scale well, \ie, far from linear speed-up according to the number of devices in use, even with state-of-the-art communication methods~\cite{sergeev2018horovod,jiang2020unified}.



The causes of low distributed training efficiency are diverse: stragglers~\cite{chen2020elastic}, computation bottlenecks \cite{hu2020distributed, ho2013more}, prolonged parameter synchronization due to 
sub-optimal tensor granularity 
\cite{peng2019generic}, 
large idling time due to poor communication-computation overlap \cite{narayanan2019pipedream}, etc. 
Effectively diagnosing performance bottlenecks and boosting distributed training 
throughput 
have been critical for the productivity of AI systems. 

Diagnosing and improving distributed training efficiency are challenging as: 
 (i) causes of unexpected performance are complex and it requires substantial time and efforts from domain experts to manually inspect runtime traces in order to figure out the real culprit; 
(ii) often, traces collected from different ML frameworks (\eg, TensorFlow \cite{abadi2016tensorflow}, PyTorch \cite{paszke2019pytorch}) are insufficient for obtaining an exact global view of a distributed training system, 
due to less accurate communication profiling;
(iii) various optimization strategies exist that can be applied to tackle performance issues, incurring a large strategy space. 

Optimization techniques for DNN training acceleration can be divided into three categories: 
1) Computation-optimization strategies, such as operator fusion \cite{jia2019optimizing, jia2019taso} and mixed precision training \cite{micikevicius2018mixed, nvidia2020mixed}; 2) Communication-oriented techniques, \ie, tensor fusion~\cite{horovod2020tensorfusion}, tensor partition \cite{peng2019generic, jiang2020unified}, gradient compression \cite{alistarh2017qsgd, 
zheng2019communication}, and transmission scheduling to improve communication-computation overlap \cite{peng2019generic,
cho2019blueconnect};
 3) Memory-optimization methods, e.g., gradient accumulation and re-computation \cite{chen2016training}.
Even within a single optimization technique, multiple possible configurations can be applied, \eg, various combinations of fusing two or more 
operators (ops) \cite{jia2019optimizing,jia2019taso}.
Besides, effects of different optimizations interact when applied together, while the combined effects 
have not been clearly explored so far.

A common approach for training performance inspection and improvement \cite{
curtsinger2015coz, 
ousterhout2015making} is to: (i) profile a training job during runtime, collecting traces which record timestamps of specific tasks and resource utilization; (ii) break down collected training time into 
computation, communication, etc., and seek insights from the summaries; (iii) apply a particular optimization technique accordingly and tune its configurations based on simulated performance or by experiments.
However, existing endeavors are limited in distributed DNN training analysis, as follows:

$\triangleright$ {\em No global data flow graph (DFG) is readily available.} 
Though local DFG can be constructed based on each worker's trace, building a global DFG including all computation and communication ops, detailed inter-worker communication topology and op dependencies is non-trivial. 
Traces independently collected from workers must be carefully aligned in execution time, 
and cross-worker communication topology, order and send-recv mapping
have to be correctly figured out from the disparate traces. 


$\triangleright$ {\em No automatic and systematic analytical tool to identify performance bottlenecks and evaluate proper optimizations. } 
Inefficiencies happen in different aspects when training different DNN models under various 
configurations, requiring different optimization strategies.  
To date, the combined effects of different optimizations and related configurations have not been carefully investigated. 

This paper proposes \sysname{}, an automatic diagnosis and optimization system for distributed DNN training expedition. 
We make the following  contributions with \sysname{}.

$\bullet$ {\sysname{}'s profiler automatically collects global traces and constructs an accurate global DFG for distributed training analysis}. 
 Computation/communication op dependencies are carefully obtained, 
 and fine-grained communication ops are exploited to model each tensor 
 transmission.   
To tackle 
clock difference among different servers and inaccurate RECV op timestamp, we carefully design a method to align trace timestamps for distributed training, exploiting dependencies between communication ops and time similarity of transmitting tensors of the same size. 

 
$\bullet$ {\sysname{}'s 
optimization framework automatically discerns performance bottlenecks and 
identifies suitable strategies for uplifting training performance}. We theoretically analyze the interactions between optimization techniques, especially op fusion and tensor fusion, and propose a new abstraction of the global DFG, \ie, the {\em \layerview}, to reduce the strategy search space. The 
algorithm framework further exploits partial replay, the critical path and the symmetry of the global DFG to accelerate strategy search. 


$\bullet$ We build the \sysname{} toolkit and release it on GitHub \footnote{\url{https://github.com/joapolarbear/dpro}}. \sysname{} can be easily enabled by setting an environment variable, and its profiling incurs little overhead.  
We evaluate \sysname{} on TensorFlow and MXNet, with PS or AllReduce gradient synchronization, using RDMA or TCP transports 
and show that \sysname{} accurately simulates distributed DNN training with a {$<5\%$} average error ($10 \times$ better than Daydream). Compared to representative baselines, dPRO effectively explores good optimization strategies, increases the training throughput by up to {$3.48 \times$} and achieves the best scalability. Finally, dPRO's search acceleration techniques can reduce the search time by orders of magnitude compared to the baselines.

\section{Background and Motivation}
\label{sec:motivation}
\subsection{DNN Training Profilers}



\mypara{1) Hardware profiling tools.} 
NVIDIA provides NVProf \cite{nvprof} to collect start/end time of each kernel, GPU core 
and memory usage, as well as many other hardware counters. NVIDIA 
CUPTI \cite{CUPTI} enables collecting profiling information at runtime. The vendor-provided tools are hardware-dependent and do not capture dependencies among ops in a DNN model, making it challenging to parse kernel-level traces. 

\mypara{2) Built-in profilers of ML frameworks.} State-of-the-art ML frameworks, such as TensorFlow \cite{abadi2016tensorflow}, PyTorch \cite{paszke2019pytorch} and MXNet \cite{chen2015mxnet}, provide their own built-in profilers. These profilers collect DNN op-level traces, including time and memory consumption when executing 
each op. 
TensorFlow and MXNet profilers also gather coarse-grained communication traces for their distributed APIs, including start time and end time of communication ops. We can not get the real transmission time as the profiling does not exclude queuing time in communication libraries.

\mypara{3) Communication library profilers.} 
Two parameter synchronization (aka communication) schemes are widely adopted for data-parallel training:
(1) AllReduce~\cite{horovod2020tensorfusion}, where all workers are connected in a tree or ring topology \cite{patarasuk2007bandwidth}, synchronously aggregate gradients using collective primitives and then update parameters independently; 
(2) parameter server (PS) architecture~\cite{jiang2020unified}, where workers \emph{PUSH} the local gradients to PSs and then \textit{PULL} the aggregated gradients from the PSs.
Horovod \cite{horovodprofiler}, a popular AllReduce communication library for DNN training, regards an entire NCCL AllReduce task for a tensor as a single op and collects start time and duration for such AllReduce tasks on a single GPU (called the `coordinator' in Horovod). BytePS \cite{bytepstimeline}, a PS-based communication library that allows tensor partition, profiles the time spent on the PUSH/PULL operation of each tensor. Their profilers do not capture computation traces. 
KungFu's profiler \cite{mai2020kungfu} is able to monitor gradient noise scale (GNS) by inserting a monitoring op to the DFG and uses a collective communication layer to aggregate GNS across workers; however, it does not track execution time of computation/communication ops and the dependencies between them.


\subsection{Challenges in Building Accurate Global Timeline}


{\em First}, existing studies (\eg, Daydream \cite{zhu2020daydream}, FlexFlow \cite{jia2018beyond}) predict training time of distributed DNN training jobs based on coarse-grained communication traces from the above existing profilers, estimating communication time based on bandwidth and tensor size. Such coarse-grained traces 
regard synchronization of one tensor as a black box without differentiating queueing time and transmission time, and are insufficient for accurately predicting distributed runtime. 
Fig.~\ref{fig:motivation} shows the per-iteration time estimated using Daydream's simulator and obtained using testbed experiments by training ResNet50~\cite{he2016deep} 
under four different configurations (using Horovod or BytePS for gradient synchronization over RDMA or TCP). 
Daydream's results remain similar across four cases, while real execution time is vastly different (due to communication protocol, topology and scale).  

{\em Next}, existing profilers do not support timestamp alignment among workers/PSs.
The error of trace timestamps of distributed training jobs is incurred by two factors: 1) there exist millisecond level or sub-millisecond level clock drifts among workers/PSs even with clock synchronization tools, such as
NTP \cite{mills1991internet} or more precise tools (\eg, HUYGENS~\cite{geng2018exploiting}), leading to some cross-worker event dependency conflicts;
2) Profiling tools can only capture the launch time of a RECV op instead of the exact time of receiving data. It is important to correct the start timestamps of RECV ops for faithful trace replay. Without trace timestamp alignment, the error of communication timestamps will accumulate and increase the error of end-to-end performance estimation.

We seek to design a generic distributed training profiler, collecting both computation and fine-grained communication traces, and aligning traces timestamps from different devices 
to provide an accurate global timeline of distributed DNN training.

\vspace{-2mm}
\subsection{DNN Training Optimization}

\mypara{Computation Acceleration.}
Op fusion \cite{tensorflowOPFS, jia2019optimizing, jia2019taso} allows a compilation of multiple computation ops to one monolithic op, achieving reduced op scheduling overhead and increased temporal and spatial locality of intermediate data access.
Mixed precision training \cite{micikevicius2018mixed, nvidia2020mixed} advocates using 16-bit floating-point instead of 32-bit as numerical formats of most ops  
for less memory consumption and faster computation. 

\mypara{Communication optimization.}
Tensor fusion \cite{horovod2020tensorfusion} fuses multiple small tensors to reduce communication overhead in gradient synchronization.
Tensor partitioning \cite{jayarajan2019priority, peng2019generic} slices a large tensor into multiple pieces to better overlap push and pull in a PS architecture.
Gradient compression \cite{alistarh2017qsgd,bernstein2018signsgd} reduces gradient size with various compression methods. 
A number of studies \cite{peng2019generic, bao2020preemptive, zhang2017poseidon} have proposed algorithms of tensor transmission scheduling to better overlap computation and communication.

\mypara{Memory optimization.} 
Re-computation \cite{chen2016training} reduces memory footprint by deleting intermediate results and re-computing them when necessary. Gradient accumulation 
accumulates gradients over consecutive 
training iterations and reduces each iteration's batch size to achieve the same overall batch size.

Choosing appropriate optimizations for a particular distributed training job is challenging, due mainly to: 

\begin{figure}[t]
\includegraphics[width=0.42\textwidth, trim = 0 0 0 0, clip]{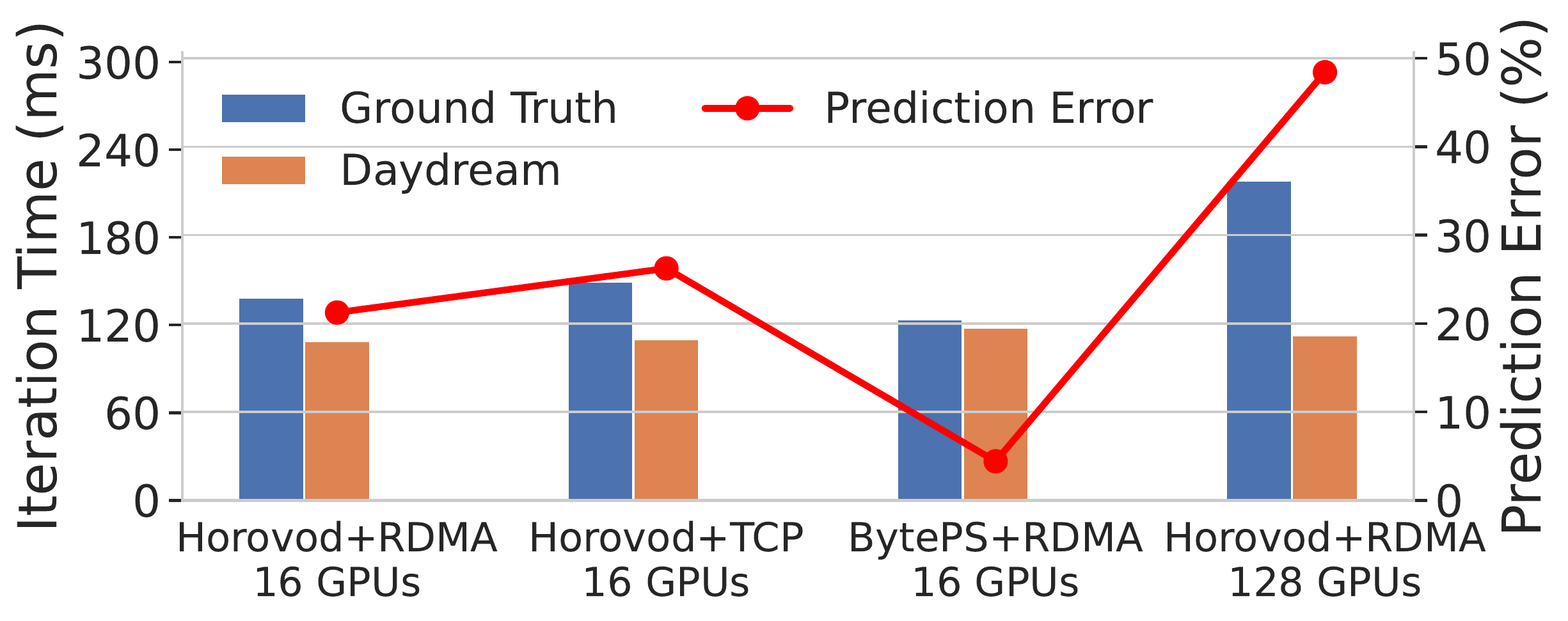}
\centering
\vspace{-2mm}
\caption{\small{Training ResNet50 in 100Gbps network, batch size 32 per GPU (see Sec.~\ref{subsec:experiment_setup} for testbed details)
}  
}
\label{fig:motivation}
\end{figure}

\begin{figure}[t]
\subfigure[]{     
\centering
\includegraphics[width=0.19 \textwidth, trim = 0 0 0 0, clip]{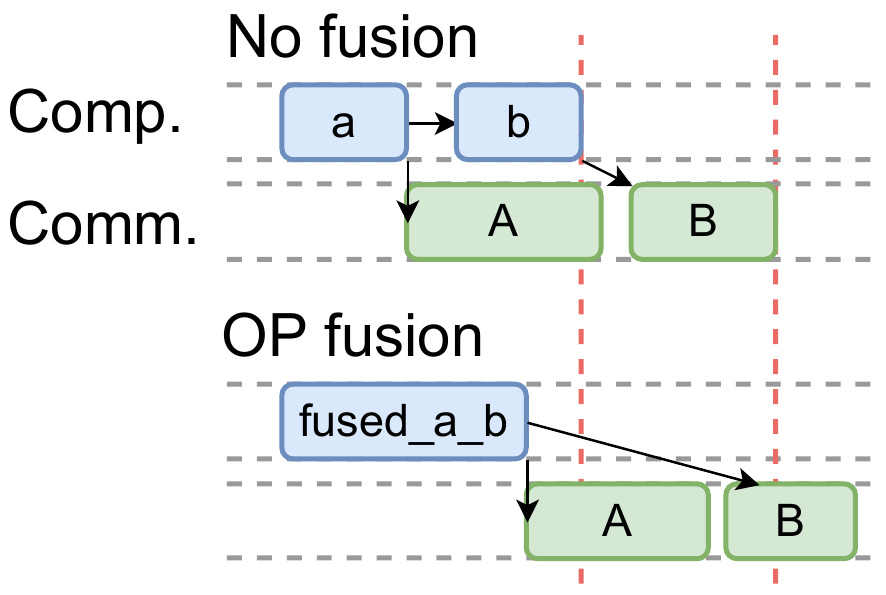}
\label{fig:motivation_opfs}
}
\subfigure[]{     
\centering
\includegraphics[width=0.255 \textwidth, trim = 0 0 0 0, clip]{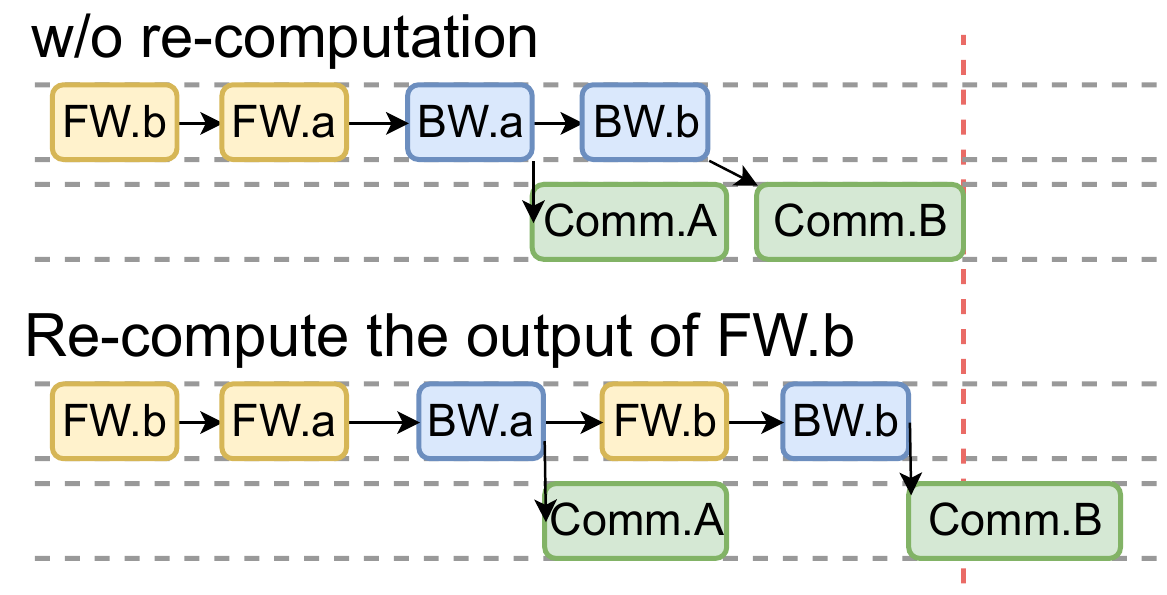}
\label{fig:motivation_recompute}
}
\vspace{-2mm}
\caption{\small{(a) Effect of op fusion: 
Comp. - computation op, Comm. - gradient synchronization; A/B is the gradient produced by op a/b; c is op fused from a and b.
(b) Effect of re-computation: FW - forward propagation; BW - backward propagation; re-computation inserts a $FW.b$ before $BW.b$ since the output of the first $FW.b$ is not cached. }}
\end{figure}

$\bullet$ {\em Interactions/conflicts among different optimizations' effects.}  
For example, DL compilers such as XLA \cite{tensorflowxla} adopt op fusion to reduce GPU memory access and may fuse all back-propagation ops (\ie, as with 
the XLA auto-clustering algorithm), which delays communication of tensors produced by these ops. As shown in Fig.~\ref{fig:motivation_opfs}, although the computation time of the fused op becomes shorter, 
end-to-end training time increases due to less overlapping between computation and communication.
Memory optimizations also interact with computation and communication. Fig.~\ref{fig:motivation_recompute} shows that re-computation of intermediate results sacrifices training speed to reduce memory footprint. 
It may also delay tensor communication. 


 

$\bullet$ {\em Very large combined strategy space.} 
The global DFG in distributed training of a 
DNN model is large, making it time-consuming to find the optimal set of strategies. 



We design a search-based automatic optimizer framework, 
that effectively explores trade-offs 
among multiple optimizations to identify proper strategies for training acceleration.  
\section{Overview}
\label{sec:overview}


\begin{figure}[t]
\includegraphics[width=0.48\textwidth]{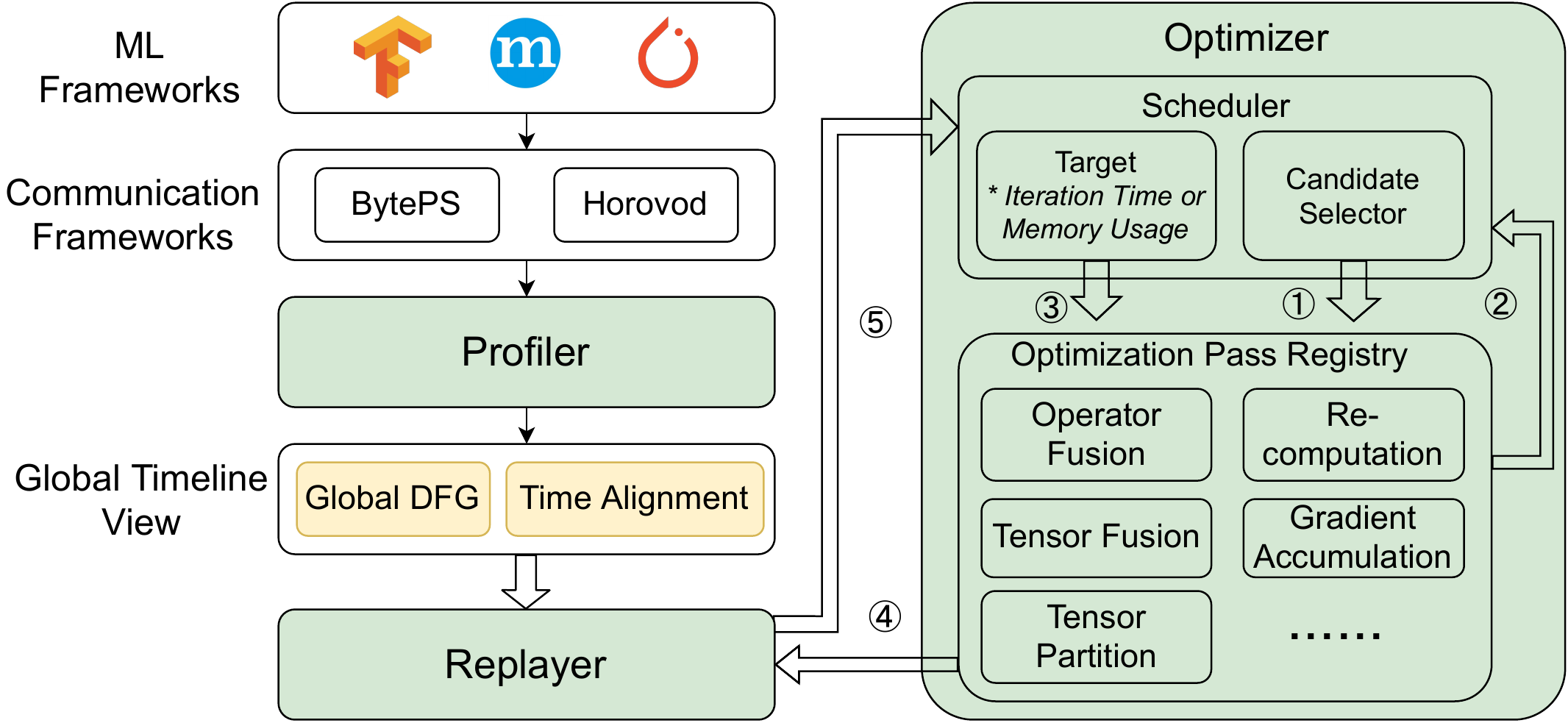}
\centering
\vspace{-2mm}
\caption{\small{\sysname{} architecture.} 
}
\label{fig:architecture}
\end{figure}


\sysname{}~includes three modules, as shown in Fig.~\ref{fig:architecture}.

\mypara{Profiler}
is a cross-framework distributed profiler, supporting three representative ML frameworks (TensorFlow, PyTorch and MXNet) and two parameter synchronization schemes (AllReduce and PS). The profiler collects time stamps (namely \textit{gTrace}) of both computation ops and fine-grained communication ops. It also tracks dependencies among fine-grained communication ops and constructs the global data flow graph (global DFG). 
Our profiler uses op-level traces for computation ops, achieving similar high simulation accuracy 
as in Daydream (which uses kernel-level traces).

\mypara{Replayer} 
simulates distributed training of 
a DNN model and estimates per-iteration training time of the global DFG.

\mypara{Optimizer} 
takes as input a given DNN model and resource configurations (\ie, GPUs and their inter-connections), 
 evaluates training performance of various strategy combinations using the replayer, 
and produces the best set 
of strategies found. 
We also provide an interface for developers to register custom 
optimization strategies. 

We detail our design of key modules in next
sections.

\section{Profiler and Replayer}
\label{sec:profiler}

\vspace{-1mm}
\subsection{Global DFG Construction}
\label{sec:global_dfg}
\label{sec:comm_profiler}



\begin{figure}[!t]
\centering
\includegraphics[width=0.45\textwidth, trim=0 0 0 0, clip]{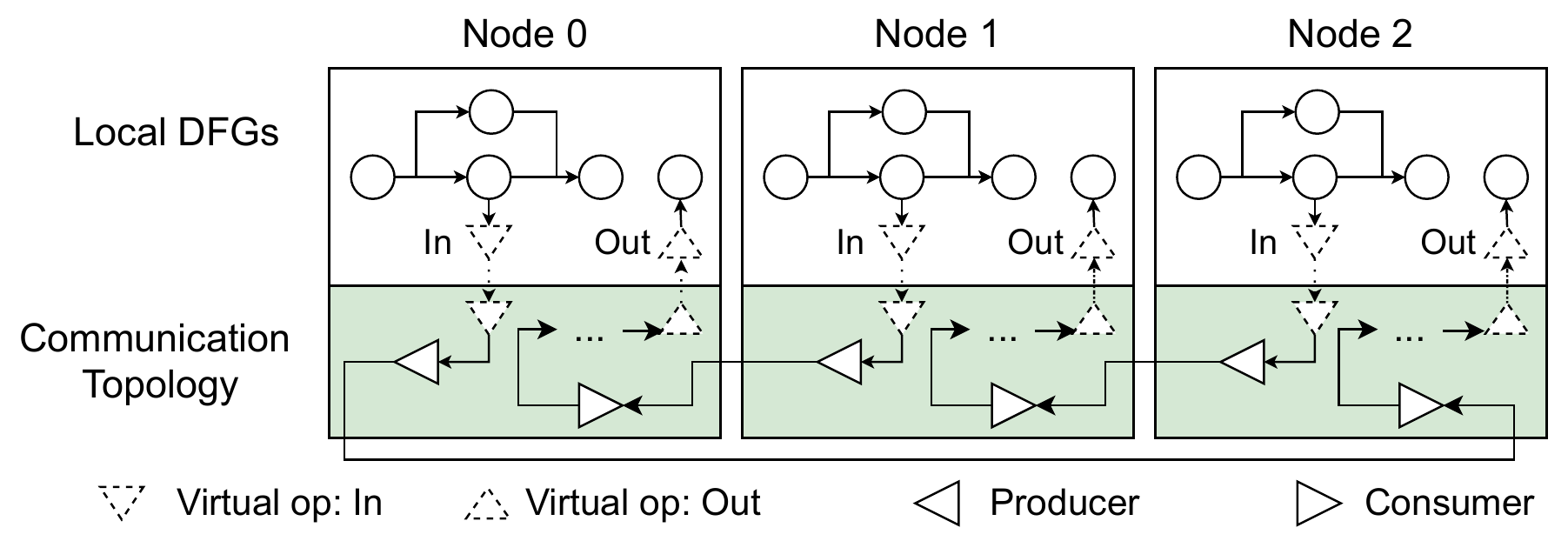}
\vspace{-2mm}
\caption{\small{An illustration of the global DFG. 
}}
\label{fig:global_dfg}
\end{figure}

The profiler automatically constructs a global DFG 
of the distributed training job, 
in which vertexes are computation and \textit{fine-grained} communication ops and edges 
denote their dependencies. 
As shown in Fig.~\ref{fig:global_dfg}, 
the global DFG consists of local data flow graphs and communication topologies.

\mypara{Local DFGs.} 
Most DL frameworks have the conception of data flow graphs for individual workers~\cite{abadi2016tensorflow, paszke2019pytorch, chen2015mxnet}. Our profiler extracts dependency information from each framework's built-in graph definition and constructs a data flow graph for each worker accordingly. We further insert a pair of In/Out virtual ops for each tensor into each local DFG, indicating where the tensor transmission occurs.
\mypara{Fine-grained Communication Topology} 
describes how each tensor is transferred between two devices, which contains two types of vertices (communication ops):
(1) \emph{producer}, which sends a tensor (partition) to another device; (2) \emph{consumer}, which receives a tensor (partition) from another device. The profiler labels every transmission of a tensor (partition) between two devices with a unique \emph{transaction ID}. 
A \emph{Middleman} connects producers to the corresponding consumers that share the same \emph{transaction ID}s. The communication topology for each tensor also contains a pair of In/Out virtual ops labeled with the tensor name, indicating the start/end of tensor transmission.

We connect local DFGs with the communication topology through In/Out virtual ops, and the global DFG is hence constructed.
Decoupling global DFG construction into local DFGs and communication topology, we enable \sysname{} to support various ML frameworks and communication architectures. In a PS architecture, each PUSH (PULL) is regarded as a pair of SEND and RECV ops at a worker (PS) and a PS (worker), respectively. The unique \emph{transaction ID} of each transmission can be produced using 
sender/receiver IPs, tensor name and whether it is PUSH or PULL. For AllReduce, we use a pair of SEND and RECV ops to represent the transmission of a chunk of the tensor to the next hop (\eg, along the ring in ring AllReduce). The \emph{transaction ID} generation further uses 
 chunk id of tensor partition and step id as in ring AllReduce \cite{ringallreduce}.

\subsection{Trace Time Alignment}
\label{clock_sync}

To combine traces collected on disparate workers/PSs and produce correct global DFG, one major obstacle is the time shift among machines where workers run. 
Not relying on accurate clock synchronization among machines, our profiler corrects the start time of 
RECV ops and obtains a more accurate communication duration of each tensor. 

Let $\mathbf{W}$ 
and $\mathbf{P}$ 
denote the set of workers and PSs in a training job, respectively. For AllReduce, $\mathbf{P}$ is empty.
Let $\bar{T}^{i}_{op}[st]$ and $\bar{T}^{i}_{op}[ed]$ be an op's measured start and end timestamps on node $i\in \mathbf{P} \cup \mathbf{W}$, and $T^{i}_{op}[st]$ and $T^{i}_{op}[st]$ represent the respective adjusted time. 
Let node $0$ be a reference for other nodes to align their time to. We compute a time offset/drift of node $i$, $\theta_i$, as the difference in measured time between node $i$ and node $0$, \ie, ${T}^{0}_{op} = \bar{T}^{0}_{op}$ and ${T}^{i}_{op} = \theta_i + \bar{T}^{i}_{op}$. We leverage two observations for time alignment.    


{\em First}, RECV ops on the same node receiving (even partitions of) the same tensor from the same sender in different training iterations, denoted as a {\em RECV op family} $f_{recv}$, should have similar execution time. 
Consider a pair of SEND and RECV from node $i$ to node $j$. Because RECV never happens before SEND,
RECV's true start time, $T_{recv}^j[st] + \theta_j$, should be 
clipped 
by the start time of SEND, $T_{send}^i[st] + \theta_i$, which changes the communication time from $(T_{recv}^j[ed] + \theta_j) - (T_{recv}^j[st] + \theta_j)$ to $T_{recv}^j[ed] + \theta_j - max(T_{recv}^j[st] + \theta_j, T_{send}^i[st] + \theta_i)$. With time adjustment, we should minimize the variance of execution time of RECV ops in the same \textit{RECV op family}: 

\vspace{-4mm}
\begin{equation*}
\small
\begin{aligned}
O_1 =  \sum_{f_{recv}\in \mathcal{F}_{recv}} Var_{recv_j \in f_{recv}} 
\left (T_{recv}^j[ed] \right. \\
\left.  + \theta_j  - max(T_{recv}^j[st] + \theta_j, T_{send}^i[st] + \theta_i) \right)
\end{aligned}
\vspace{-2mm}
\end{equation*}

where $\mathcal{F}_{recv}$ is the set of all \textit{RECV op families} and $i$ denotes the node where the sender of the tensor of $f_{recv}$ resides. 

{\em Second}, nodes on the same physical machine should have the same time offset because they share the same physical clock. Let $\mathcal{M}$ be the set of all physical machines and $g_m$ be the set of nodes on machine $m$. We should ensure the time offset of workers on the same machine as close as possible: 

\vspace{-4mm}
\begin{equation*}
\small
\begin{aligned}
O_2 =\sum_{m \in \mathcal{M}}Var_{i\in g_m}(\theta_i)
\end{aligned}
\vspace{-2mm}
\end{equation*}

We compute time offsets $\theta_i$'s for time alignment among distributed nodes, by solving the following optimization:

\vspace{-5mm}
\begin{equation*}
\small
\begin{aligned}
~~~~~~~~~~& \min_{\theta_i: i\in \mathbf{P} \cup \mathbf{W}} a_1 O_1+a_2O2 \\
~~~~~~~~~~& ~~\textrm{s.t.} \quad \theta_0=0, \\
~~~~~~~~~~& ~~~~~~~~~~ \theta_{i} - \theta_{j} \leq \bar{T}^{j}_{o_2} - \bar{T}^{i}_{o_1},\\
~~~~~~~~~~&~~~~~~
\forall (i,j) \in (\mathbf{W} \times \mathbf{P}) \cup (\mathbf{P} \times \mathbf{W}),i \neq j, (o_1, o_2) \in \mathrm{E}
\end{aligned}
\vspace{-1mm}
\end{equation*}
Here $a_1\ge 0$ and $a_2\ge 0$ are two coefficients gauging the weights of the two objectives, and $E$ is the set of inter-op dependencies. 
The constraints ensure inter-op dependencies for time alignment, \ie, the adjusted time of an op $o_2$ on $j$ ($\bar{T}_{o_2}^j + \theta_j$) that depends on op  $o_1$ on $i$, is not earlier than the adjusted time of $o_1$ ($\bar{T}_{o_1}^i + \theta_i$). 
The optimization problem can be solved in a few seconds using the state-of-the-art optimization libraries \cite{cvxpy}. 

\vspace{-2mm}
\subsection{Replayer}
\label{subsec:replayer}
The replayer simulates the execution of the global DFG 
based on a modified Kahn’s algorithm \cite{kahn1962topological} of topological sorting. Instead of using a global ready queue (used in Daydream or Kahn's algorithm), for a distributed training job, we regard each worker/PS and each communication link as one \emph{device}, and the replayer maintains a queue and a device time (end time of the last op executed on the device) for each device. An op is enqueued to the queue of corresponding device once it is ready, i.e., all its predecessor ops are executed. The replayer iteratively picks the device with the smallest device time, dequeues an op from the head of this queue and updates the corresponding device time with the op's execution time. After all ops in the global DFG are run, we take the largest device time as the iteration time. 
Although there might be multiple feasible topological sortings, dPRO's replayer can generate the most likely one by averaging op execution time over 10 training iterations and imitating the FIFO queue in ML framework's engines.

The replayer also produces an execution graph by adding additional edges into the global DFG, indicating execution ordering between ops running in the same device, and computes the critical path on the execution graph, for bottleneck identification by the optimizer. 

\section{Optimizer}
\label{sec:optimizer}

Given a 
global DFG $\mathcal{G}$ 
of a distributed DNN training job and a set of optimization strategies $\mathcal{S}$, the optimizer identifies the bottlenecks in the global graph through training replay and produces a subset of optimization strategies, $\mathcal{S}^* \in  \mathcal{S}$, minimizing per-iteration training time (referred to as the iteration time):

\vspace{-4mm}
\begin{equation*}
\small
\begin{aligned}
min_{\mathcal{S}^\prime \in  \mathcal{S}} \textsc{IterationTime}(f(\mathcal{G}, \mathcal{S}^\prime))
\end{aligned}
\vspace{-2mm}
\end{equation*}

where $\mathcal{G}^\prime = f(\mathcal{G}, \mathcal{S}^\prime)$ is the modified global DFG after applying strategies in $\mathcal{S}^\prime$ to the original global DFG $\mathcal{G}$.

\subsection{Theory Foundation}

\begin{figure}[!t]
\centering
\includegraphics[width=0.495\textwidth, trim=0 0 0 0, clip]{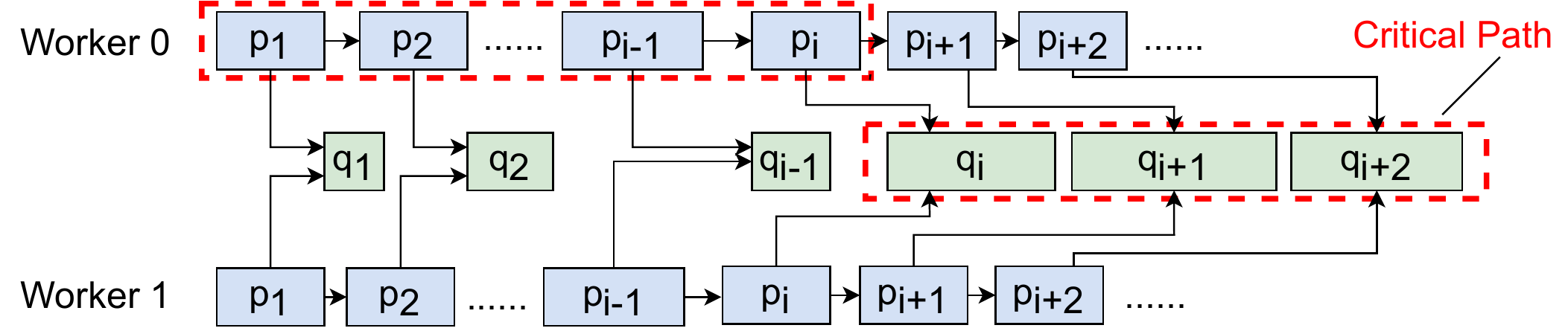}
\vspace{-2mm}
\caption{\small{Illustration of Critical Path.
}}
\label{fig:critical_path}
\vspace{-2mm}
\end{figure}

The main idea of our optimizer is to iteratively check and optimize the critical path of the global execution graph. 
The critical path $C$ contains a sequence of computation and communication ops: $C = [p_0, p_1, ..., p_{i}, q_{i}, q_{i+1}, ..., q_{|C|-1}]$, where $p_0, p_1, ..., p_{i}$ are computation ops and $q_i, q_1, ..., q_{|C|-1}$ are communication ops. Here, we group fine-grained communication ops in the global DFG for the transmission of tensor $n$ (e.g., SEND and RECV) as one communication op $q_n$. Fig.~\ref{fig:critical_path} depicts an example of the critical path and the correspondence between each pair of $p_n/q_n, n=0, 1, ...|C|-1$. Since 
gradient tensors are dependent on computation ops, the critical path always starts from a sequence of computation ops and ends with a sequence of communication ops.

Note that each computation op $p_n, n=1,\ldots, i$ on the critical path may correspond to a communication op $q_n$ (each backward computation op has a corresponding tensor communication op, while we treat $q_n$ for a forward computation op $p_n$ as null); the corresponding communication ops do not lie on the critical path before $p_i$, because computation ops are the bottleneck in this phase;
On the other hand, each communication op $q_n, n=i, \ldots, |C|-1$ on the critical path corresponds to a computation op $p_n$ which may not lie on the critical path. 

The optimizer inspects the critical path form $p_0$ to $q_{|C|-1}$. For each op, $p_n, n=1,\ldots,i$ or $q_n, n=i,\ldots, |C|-1$,
a decision $d_n$ is made on whether op fusion and/or tensor fusion should be applied: 
1) $d_n = opfs$, fusing two computation ops $p_{n-1}$ and $p_n$;
2) $d_n = tsfs$, fusing the two tensors 
$q_{n-1}$ and $q_n$ (those corresponding to computation ops $p_{n-1}$ and $p_n$);
3) $d_n = opfs\_tsfs$, fusing $p_{n-1}$ and $p_n$ and fusing tensors $q_{n-1}$ and $q_n$;
4) $d_n = null$, no fusion.
We have $d_0 = null$. 
When tensor partition is enabled, 
the optimizer also decides an optimal partition number $k_n$ for the tensor of op $p_n$ or $q_n$ on the critical path. Unlike op and tensor fusion, tensor partition does not hurt computation-communication overlapping, 
but only affects tensor synchronization time, which inspires us to adopt the optimal partition size for each possible choice of $d_n$, before comparing their performance.

Let $T_n$ denote the duration from the start of the global DFG execution till the completion of computation op $p_n$ and corresponding communication op $p_n$, \ie, $T_n = \max(p_n^e, q_n^e)$ (see Table~\ref{table:notation} for notation definitions). 
The optimizer finds optimal decisions $\mathcal{D} = [d_0, d_1, ..., d_{|C|-1}],$ and $\mathcal{K} = [k_0, k_1, ..., k_{|C|-1}]$ that minimize the duration of the critical path: $\min_{\mathcal{D}, \mathcal{K}} T_{|C|-1}$.



\begin{table}[!t]
    \small
    \vskip -0.05in
    \caption{\small{Notation}}
    \vskip 0.05in
    \label{table:notation}
    \begin{tabular}{|p{0.05\textwidth}<{}|p{0.2\textwidth}<{}|p{0.15\textwidth}<{}|} 
        \hline
        Symbol & Description & Calculation \\
        \hline
        $p_n^d$ & execution time of computation op $p_n$ & profiled 
        \\
        \hline
        $p_n^e 
        $ & end 
        time of computation op $p_n$ & decided during the search process \\
        \hline
        $q_n^d$ & execution time of synchronizing tensor $q_n$ & estimated using partial replay \\ 
        \hline 
        $q_n^s$ & size of tensor $q_n$ & profiled \\
        \hline
        $q_n^e 
        $ & end 
        time of communication op $q_n$ 
        & decided during the search process  \\  
        \hline 
        $d_n$ & strategy taken on the $n$-th op in the critical path & decided during the search process \\
        \hline
        $k_n$ & partition number of tensor $q_n$ & computed during the search process \\
        \hline
    \end{tabular}
\end{table}

We analyze conditions for applying 
the optimization techniques, 
deriving insights to tailor a strategy search algorithm.
Let $opfs\_time(p_{n-1}, p_n)$ be the execution time of the fused op after fusing $p_{n-1}$ and $p_n$. 

\vspace{-2mm}
\begin{theorem}[Op Fusion]\label{theorem:opfs}
If $q_{n-1}^d \leq p_n^d + p_{n-1}^d - opfs\_time(p_{n-1}, p_n)$, $T_n$ achieved with this op fusion is smaller than no fusion, \ie, $T_n(d_n=opfs) \leq T_n(d_n=null)$ 
(fusing $p_{n-1}$ and $p_n$ is better than not); otherwise, 
$T_n(d_n=opfs) \geq T_n(d_n=null)$, 
 \ie, fusing $p_{n-1}$ and $p_n$ leads to a worse performance.
\end{theorem}
\vspace{-2mm}


Let $t_{sync}(s, k) $ be the time to synchronize a tensor of size $s$, divided to $k$ partitions, \ie, execution time of the complete synchronization operation on this tensor (using either PS or AllReduce). 
Given a tensor of size $s$, we use $k^*[s]$ to denote the optimal partition number that minimizes $t_{sync}$. 
\vspace{-2mm}
\begin{theorem}[Tensor Fusion/Partition]\label{theorem:tsfs}
If $q_{n-1}^e > p_n^e + t_{sync}(q_{n-1}^s + q_n^s, k^*[q_{n-1}^s + q_n^s]) - t_{sync}(q_n^s, k^*[q_n^s])$, then $T_n$ achieved by fusing tensors $q_{n-1}$ and $q_n$ is smaller than no fusion, \ie, $T_n(d_n=tsfs, k_n=k^*[q_{n-1}^s + q_n^s])<T_n(d_n=null, k_n=k^*[q_n^s])$ (fusing $q_{n-1}$ and $q_n$ is better than not); otherwise, $q_{n-1}$ and $q_n$ should not be fused.
\end{theorem}
\vspace{-2mm}

When considering op fusion and tensor fusion/partition together, 
if fusing two computation (communication) ops is better than not, their corresponding communication (computation) ops - if there are - should also be fused, 
without sacrificing computation-communication overlapping.

\vspace{-2mm}
\begin{theorem}[Op Fusion and Tensor Fusion/Partition]\label{theorem:opfs_tsfs}
$T_n(d_n=opfs\_tsfs, k_n=k^*[q_{n-1}^s + q_n^s]) \leq T_n(d_n=tsfs, k_n=k^*[q_{n-1}^s + q_n^s])$ and $T_n(d_n=opfs\_tsfs, k_n=k^*[q_{n-1}^s + q_n^s]) \leq T_n(d_n=opfs, k_n^*[q_n^s])$.
\end{theorem}
\vspace{-2mm}

See the appendix \cite{hu2022appendix} for proofs of Theorems \ref{theorem:opfs}, \ref{theorem:tsfs} and \ref{theorem:opfs_tsfs}.

\begin{figure}[!t]
\centering
\includegraphics[width=0.495\textwidth, trim=0 0 0 0, clip]{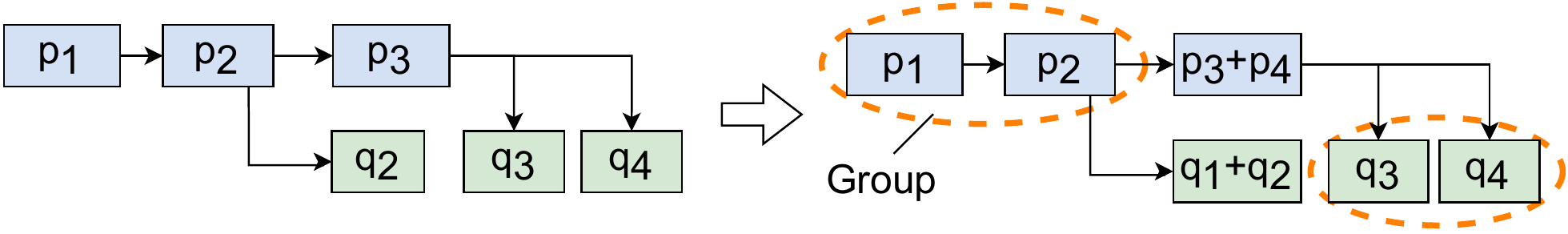}
\vspace{-2mm}
\caption{\small{Illustration of \layerview, where $p_1$ has no learnable parameter but $p_3$ has two.
}}
\vspace{-2mm}
\label{fig:layerview}
\end{figure}

\subsection{Diagnosis and Optimization Algorithm}
An ablation of \sysname{}'s optimizer module is given in Fig.~\ref{fig:architecture}. The optimizer maintains a \textit{Graph Pass Registry} including various optimization techniques. Each \textit{Graph Pass} in the Registry corresponds to an optimization technique, such as op fusion, tensor fusion, tensor partition, 
etc. 

The optimizer analyzes the bottlenecks in the global DFG and optimizes them in an iterative manner. The optimizer algorithm is given in Alg.~\ref{algo_search}. At the beginning, the optimizer evaluates the iteration time and memory usage of the original global DFG $\mathcal{G}$ by replaying it with the replayer. If the memory usage exceeds the memory budget (specified by the user), the memory-optimization passes are invoked to reduce the memory footprint (see the appendix \cite{hu2022appendix} for details).


Then the optimizer proceeds with throughput optimization, that minimizes the makespan of the critical path in the global execution graph. 
Given a critical path $C = [p_0, p_1, ..., p_i, q_i, q_{i+1}, ..., q_{|C|-1}]$, 
  the optimizer first examines the computation ops $p_n,n=1,\ldots,i$ in order. 
  Because the performance of this segment of the critical path is computation-bound, the optimizer first evaluates whether op fusion should be applied on $p_{n-1}$ and $p_n$ according to Theorem \ref{theorem:opfs}. If 
  so, it invokes the op fusion pass to fuse $p_{n-1}$ and $p_n$, 
  as well as the tensor fusion pass to fuse the corresponding two tensors $q_{n-1}$ and $q_n$ (Theorem \ref{theorem:opfs_tsfs}). An optimal partition number $k^*$ is then computed and applied (if $k^*>1$) on the fused or non-fused tensor $q_n$.

Next, the optimizer inspects the communication ops $q_n, n=i,\ldots,|C|-1$, on the critical path. In this segment, 
the performance is communication-bound. 
The optimizer first computes the optimal partition number $k^*$ on the fused and non-fused tensor $q_n$ and checks whether 
tensor fusion should be applied to 
$q_{n-1}$ and $q_n$ according to Theorem \ref{theorem:tsfs}. If 
so, the optimizer invokes the tensor fusion pass to fuses $q_{n-1}$ and $q_n$, as well as the corresponding computation ops $p_{n-1}$ and $p_n$ (Theorem \ref{theorem:opfs_tsfs}). 
Then the optimal partition number $k^*$ on fused or non-fused tensor is applied accordingly.


After applying the optimizations, 
the global DFG $G$ is updated. The optimizer uses the replayer to update the critical path $C$ 
and then repeats the search on the new critical path. 

The fused op execution time, $opfs\_time(p_{n-1}, p_n)$, can be obtained by profiling the fused op in an offline manner (as we will do in the experiments) or use a cost model \cite{kaufman2019learned}.
The time to synchronize a tensor, $t_{sync}(s, k)$, is estimated with partial replay of the subgraph including all relevant communication ops 
(Sec.~\ref{sec:speedup}). 
The optimal partition number of a tensor is obtained through grid search. 


\begin{algorithm}[!t]
\caption{Diagnosis and Optimization Algorithm} 
\label{algo_search}
\small
\begin{algorithmic}[1]
\STATE If OOM, invoke the Memory Optimization Pass
\STATE Construct the {\em \layerview}~ and fuse ops/tensors in the same {\em \layername}.
\STATE The replayer computes an initial critical path $C = [p_0, p_1, ..., p_i, q_i, q_{i+1}, ..., q_{|C|-1}]$
\WHILE{search time budget not exhausted and speed-up not converged}
    \FOR{$n=0\rightarrow i$}
    	\IF{$q_{n-1}^d < p_n^d + p_{n-1}^d - opfs\_time(p_{n-1}, p_n)$}
    		\STATE \textsc{OPFusion}($p_{n-1}$, $p_n$) [Theorem \ref{theorem:opfs}]
    		\STATE \textsc{TensorFusion}($q_{n-1}$ , $q_n$) [Theorem \ref{theorem:opfs_tsfs}]
    		\STATE $k^* \leftarrow \textsc{OptPartNum}(q_{n-1}+q_n)$
    		\STATE Partition fused tensor $q_{n-1}^s + q_n^s$ evenly by $k^*$ 
  		\ELSE 
  		    \STATE // tensor partition only
            \STATE $k^* \leftarrow \textsc{OptPartNum}(q_n)$
    		\STATE Partition tensor $q_n^s$ evenly by $k^* $
    \ENDIF
    \ENDFOR
	\FOR{$n=i  \rightarrow |C|-1$}
    	\IF{$q_{n-1}^e > p_n^e + t_{sync}(q_{n-1}^s + q_n^s, k^*[q_{n-1}^s + q_n^s]) - t_{sync}(q_n^s, k^*[q_n^s])$}
    		\STATE \textsc{TensorFusion}( $q_{n-1}$, $q_n$) [Theorem \ref{theorem:tsfs}]
    		\STATE \textsc{OPFusion}( $p_{n-1}$ , $p_n$)  [Theorem \ref{theorem:opfs_tsfs}]
    		\STATE
    		Partition fused tensor $q_{n-1}+q_n$ evenly by $k^* = \textsc{OptPartNum}(q_{n-1}+q_n)$ 
    	\ELSE
    		\STATE Partition tensor $q_n^s$  evenly by $k^*=\textsc{OptPartNum}(q_n)$ 
    	\ENDIF
    \ENDFOR

  	\STATE Execute the updated global DFG using the replayer and obtain a new critical path $C = [p_0, p_1, ..., p_i, q_i, q_{i+1}, ..., q_{|C|-1}]$
\ENDWHILE
\end{algorithmic}
\end{algorithm}


\subsection{Search Speed-up} \label{sec:speedup}
It is time-consuming to exhaustively explore the large strategy space, \eg, it takes more than 24 hours to search for the optimal optimization strategies for BERT Base. 
We propose several techniques to expedite the search process.

\textbf{\layerview.}
Inspired by Theorem \ref{theorem:opfs_tsfs}, we 
coarsen the global DFG during the search process, namely constructing the \textit{\layerview}: we put computation ops that do not produce tensors but are close to one tensor-producing  computation op into one \layername~(together with the latter), and communication ops connected to the same computation op into one \layername. 
Ops or tensors in the same \layername~are fused, and we search for optimization strategies based on such a coarsened view~of the global DFG  (line 1 of Alg.~\ref{algo_search}). Fig.~\ref{fig:layerview} gives an illustration. 
$p_1$ produces no tensor ($q_1=null$) and is connected to $p_2$ which generates a tensor $q_2$; $p_1$ and $p_2$ are fused into one \layername. 
The rationale lies in that: we can view tensor $q_2$ as a fusion of $q_1=null$ and $q_2$; then according to Theorem \ref{theorem:opfs_tsfs}, fusing $p_1$ and $p_2$ leads to a better performance. 
$q_3$ and $q_4$ are both tensors produced by $p_3$ (e.g., the BatchNorm Layer has two learnable parameters \cite{ioffe2015batch}), and they are put into one \layername. 
This is because we can regard $p_3$ as a fusion of $p_3$ and $p_4=null$; then fusing $q_3$ and $q_4$ is better than not  based on Theorem \ref{theorem:opfs_tsfs}. 
We construct \textit{\layerview} using a backtracking algorithm detailed in the appendix \cite{hu2022appendix}.


\textbf{Partial Replay.}
To avoid frequently replaying the entire global DFG during strategy search (for estimating $t_{sync}(s, k)$), the replayer supports partial simulation. Given the current global DFG $G$, 
the replayer identifies all  
communication ops $\mathcal{S}_{p}$ related to the tensor, 
and generates a subgraph $G^\prime$, which contains the ops in $\mathcal{S}_{p}$ and the edges between those ops. Execution of the subgraph is simulated 
to produce the tensor synchronization time. 


\textbf{Exploiting Symmetry.} 
We further leverage the symmetry in the global DFG of state-of-the-art DNNs to accelerate strategy search. 
For example, BERT~\cite{devlin2018bert} includes multiple transformer blocks; the strategies applied inside one block can be used in other blocks as well. For data parallel training with homogeneous workers, the optimizations applied on one critical path can actually be applied to multiple workers. 

\section{Implementation}


\mypara{Profiler.} To profile computation ops, we use the native profiler of each ML framework with minor modifications.

$\bullet$ {\em TensorFlow.} We implemented \sysname{} with TensorFlow 2.4 in the graph execution mode. We modify 
tf.profiler to collect computation traces with absolute time and extract dependencies from \textit{tf.RunMetadata.partition\_graphs}. 

$\bullet$ {\em MXNet.}
We collect computation traces using 
$mxnet.profiler$, 
with some modifications to ensure a unique trace name for each op. 
We extract op dependencies through \textit{$mxnet.Symbol.debug\_str()$} API. 


$\bullet$ {\em Communication.}
We use Horovod \cite{sergeev2018horovod} as the 
AllReduce communication library, which uses 
NCCL \cite{nccl} for collective communication across GPUs.
We dive into NCCL (adding 318 LoCs) to collect timestamps of SEND/RECV of the tensor chunks.
We adopt BytePS \cite{jiang2020unified} for PS-based communication and add around 400 LoCs to its communication library, ps-lite \cite{li2014scaling}, for recording timestamps of PUSH and PULL of each tensor. 
\mypara{Replayer.}
We implement it 
using Python with 3653 LoCs. 

\mypara{Optimizer.} We implement the optimizer and all optimization passes in Python with 5745 LoCs.
We implement op fusion on TensorFlow using XLA \cite{tensorflowxla}, and modify it 
to allow manual control of which ops to be fused in detail, in order to apply our identified strategies. 
We modify Horovod and BytePS to enable customized tensor fusion pattern and tensor partition size.

\mypara{APIs and Commands.} \sysname{}~provides a simple programming interface for trace collection. 
The following shows the APIs for trace collection on TensorFlow, with only 2 additional LoCs: a \emph{recorder} is defined, with which the wrapper 
decides when to start/finish profiling and output traces.

\vspace{-2mm}
\begin{lstlisting}
recorder = dpro.Recorder(model)
@dpro.profile(recorder)
def train_one_step():
    # define one training step here
\end{lstlisting}
\vspace{-4mm}
After profiling, a user can call the commands as follows 
to invoke the profiler, replayer and optimizer, respectively.

\vspace{-2mm}
\begin{lstlisting}
$ dpro profile <python program> -o <trace path>
$ dpro replay <trace path> 
$ dpro optimize <trace path> 
\end{lstlisting}
\vspace{-2mm}

\begin{figure*}[!t]
\centering
\includegraphics[width=0.76\textwidth, trim=0 50 0 35, clip]{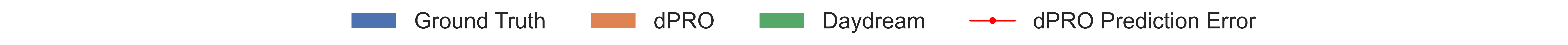}
\subfigure[ResNet50, TCP]{     
\centering
\includegraphics[width=0.235\textwidth, trim=0 0 0 0, clip]{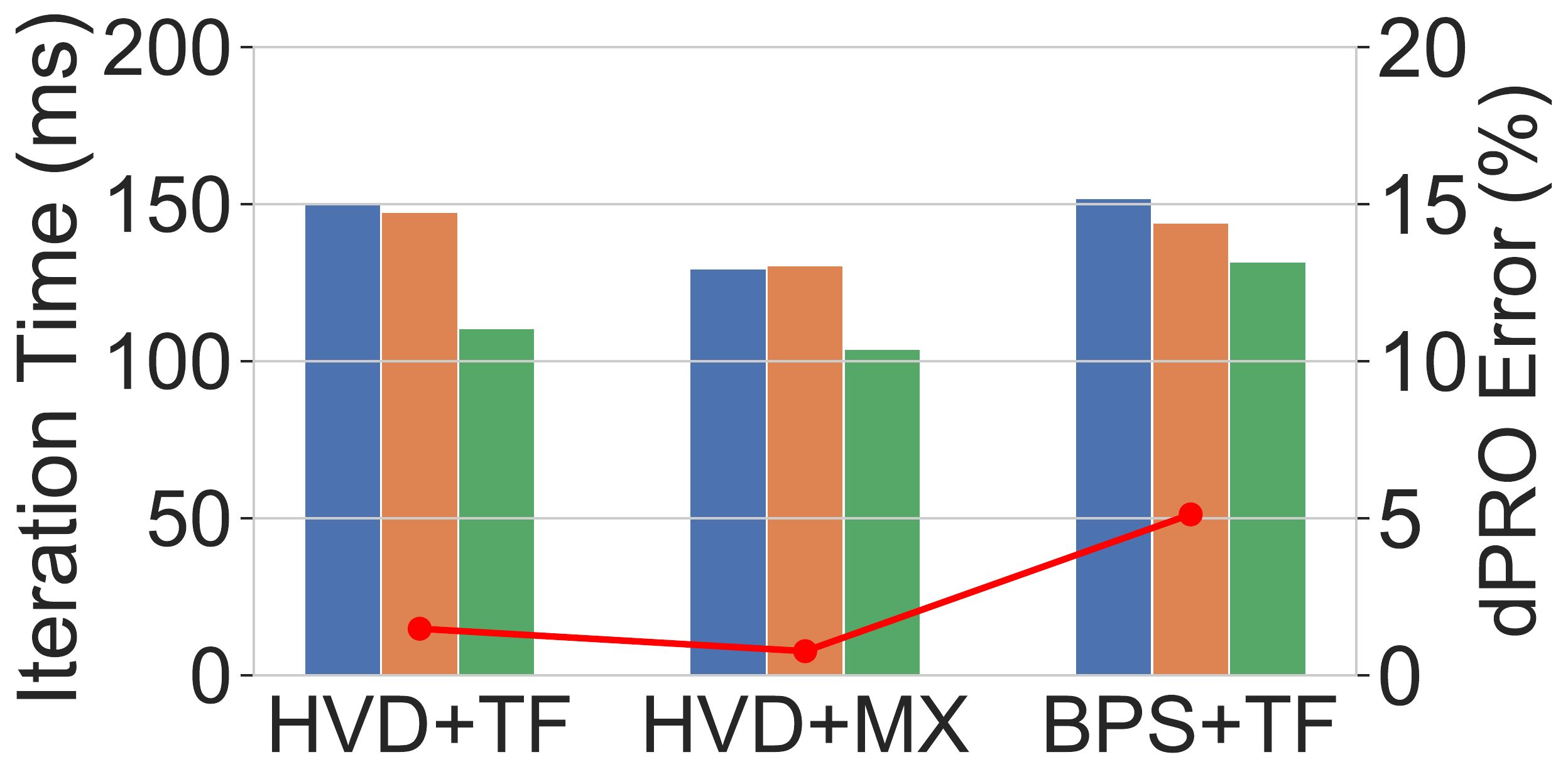}
}
\subfigure[ResNet50, RDMA]{     
\centering
\includegraphics[width=0.235\textwidth, trim=0 0 0 0, clip]{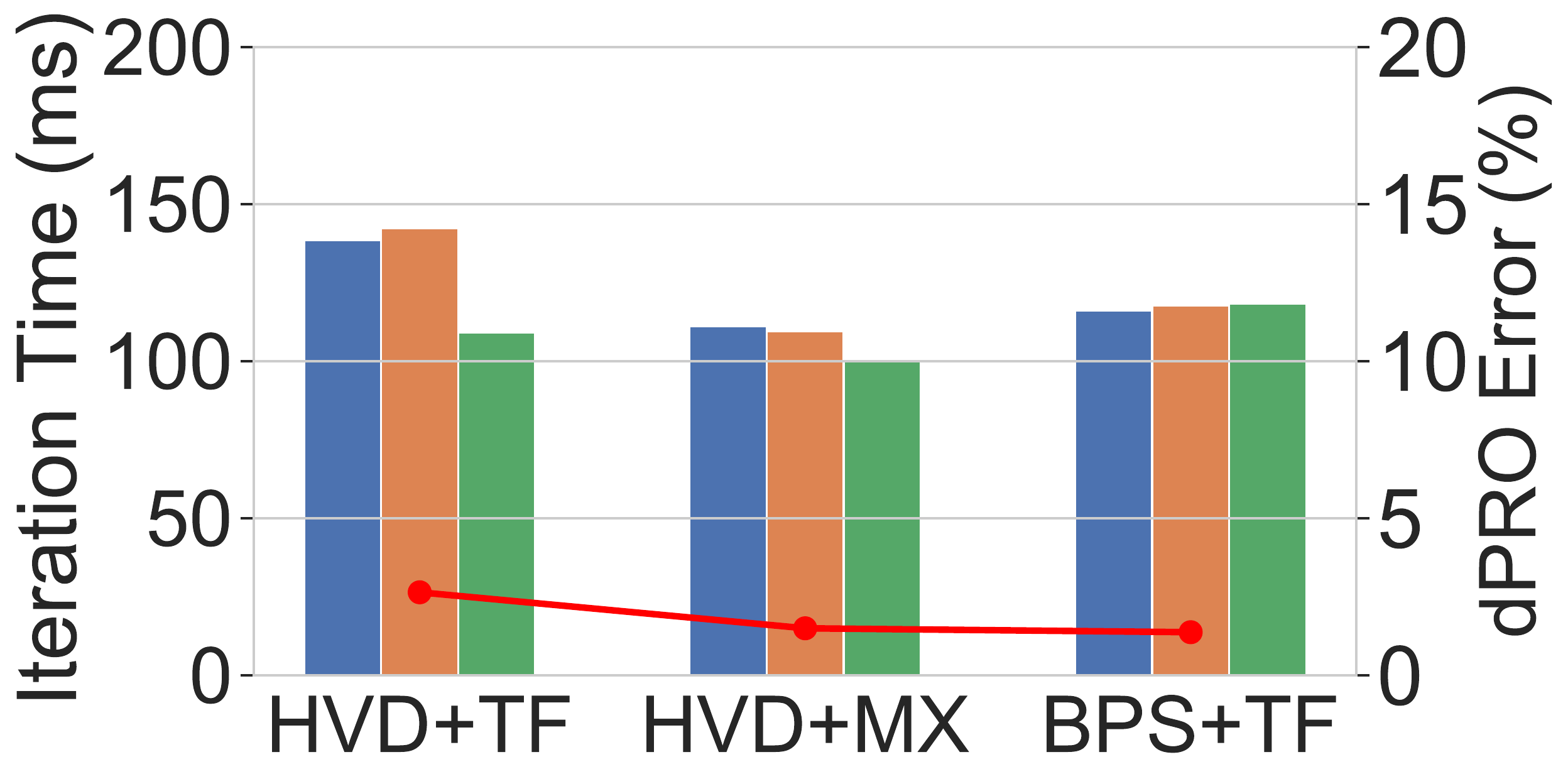}
}
\subfigure[VGG16, TCP]{     
\centering
\includegraphics[width=0.235\textwidth, trim=0 0 0 0, clip]{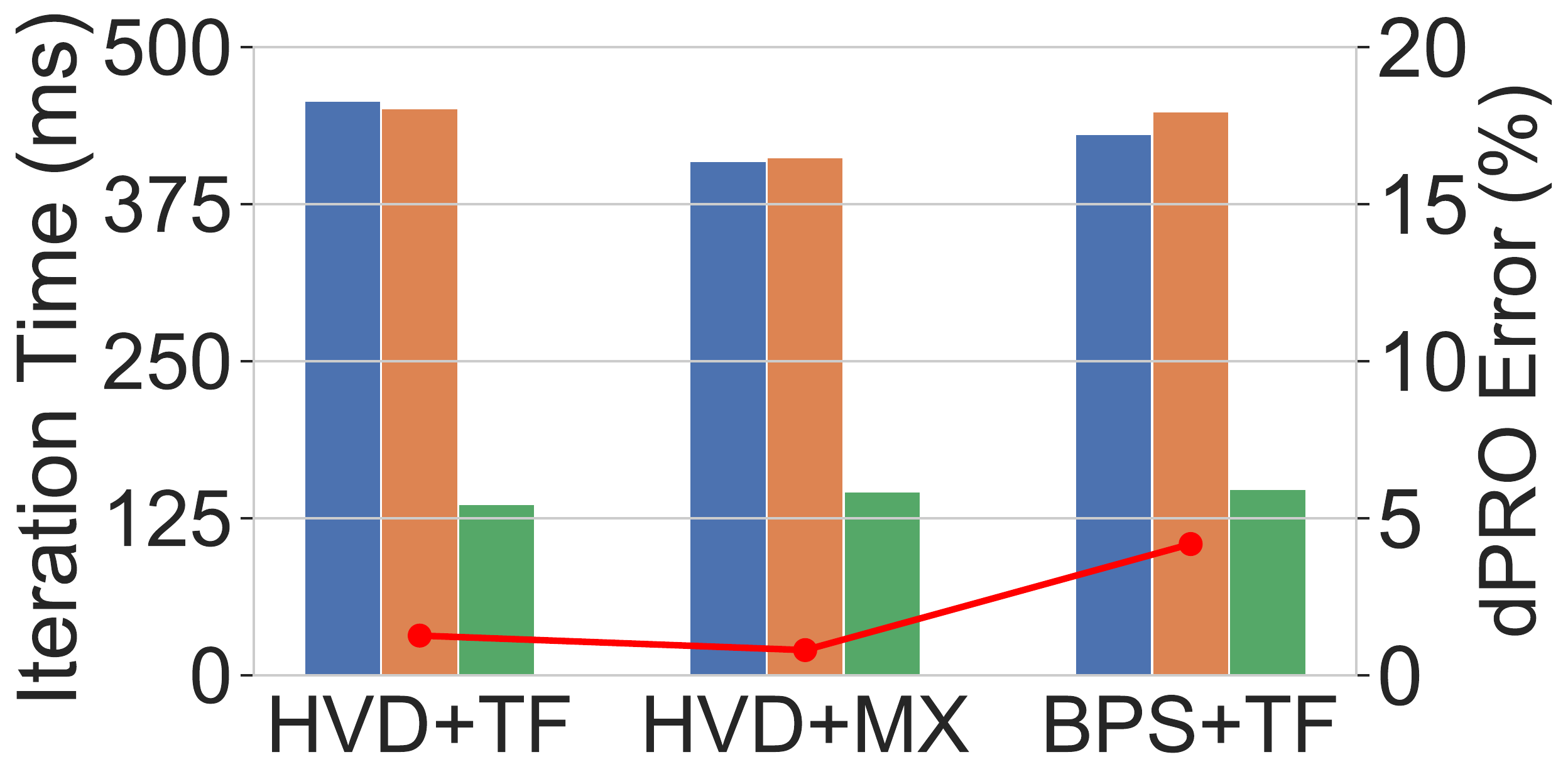}
}
\subfigure[VGG16, RDMA]{     
\centering
\includegraphics[width=0.235\textwidth, trim=0 0 0 0, clip]{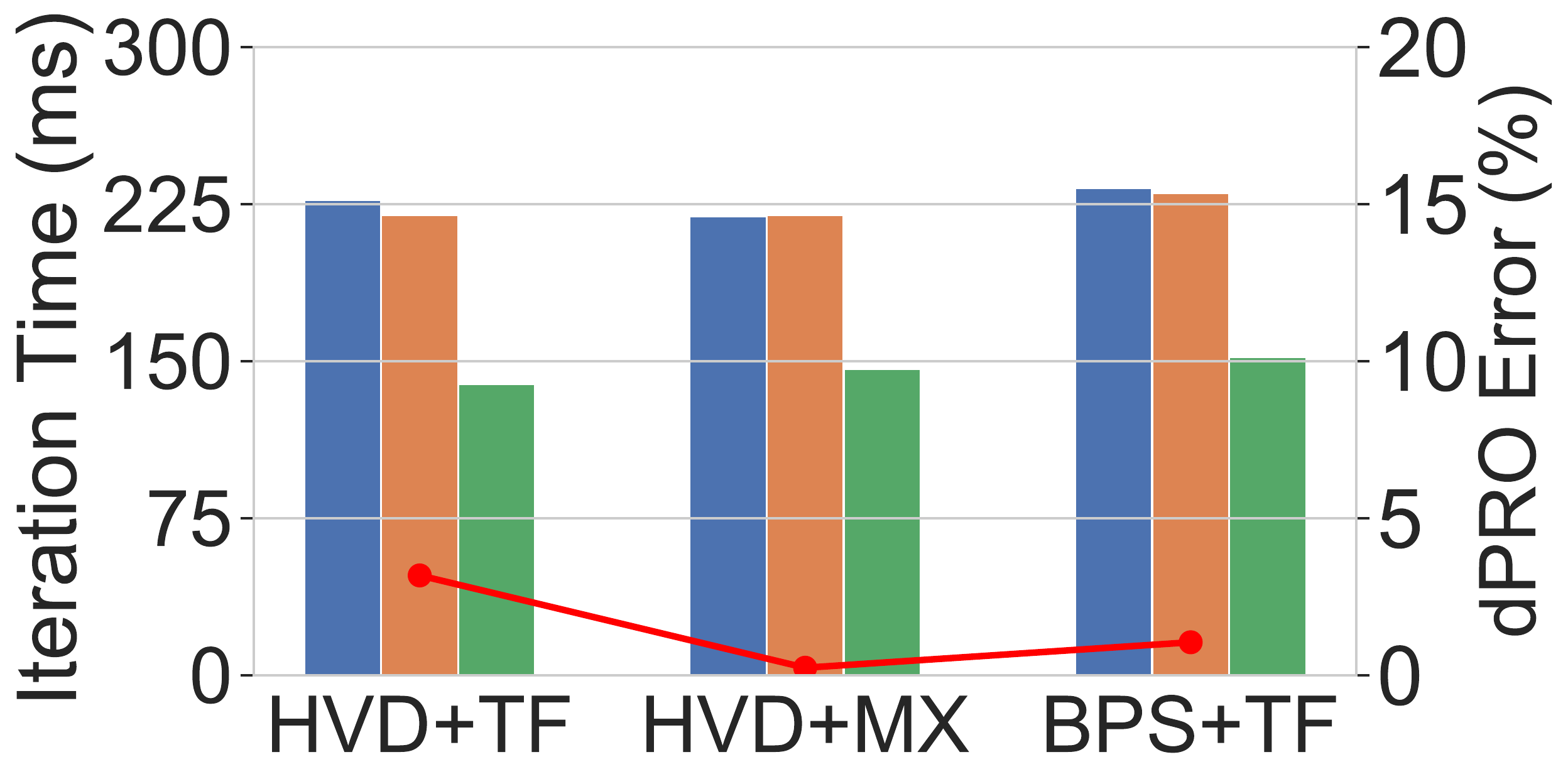}
}
\subfigure[InceptionV3, TCP]{     
\centering
\includegraphics[width=0.235\textwidth, trim=0 0 0 0, clip]{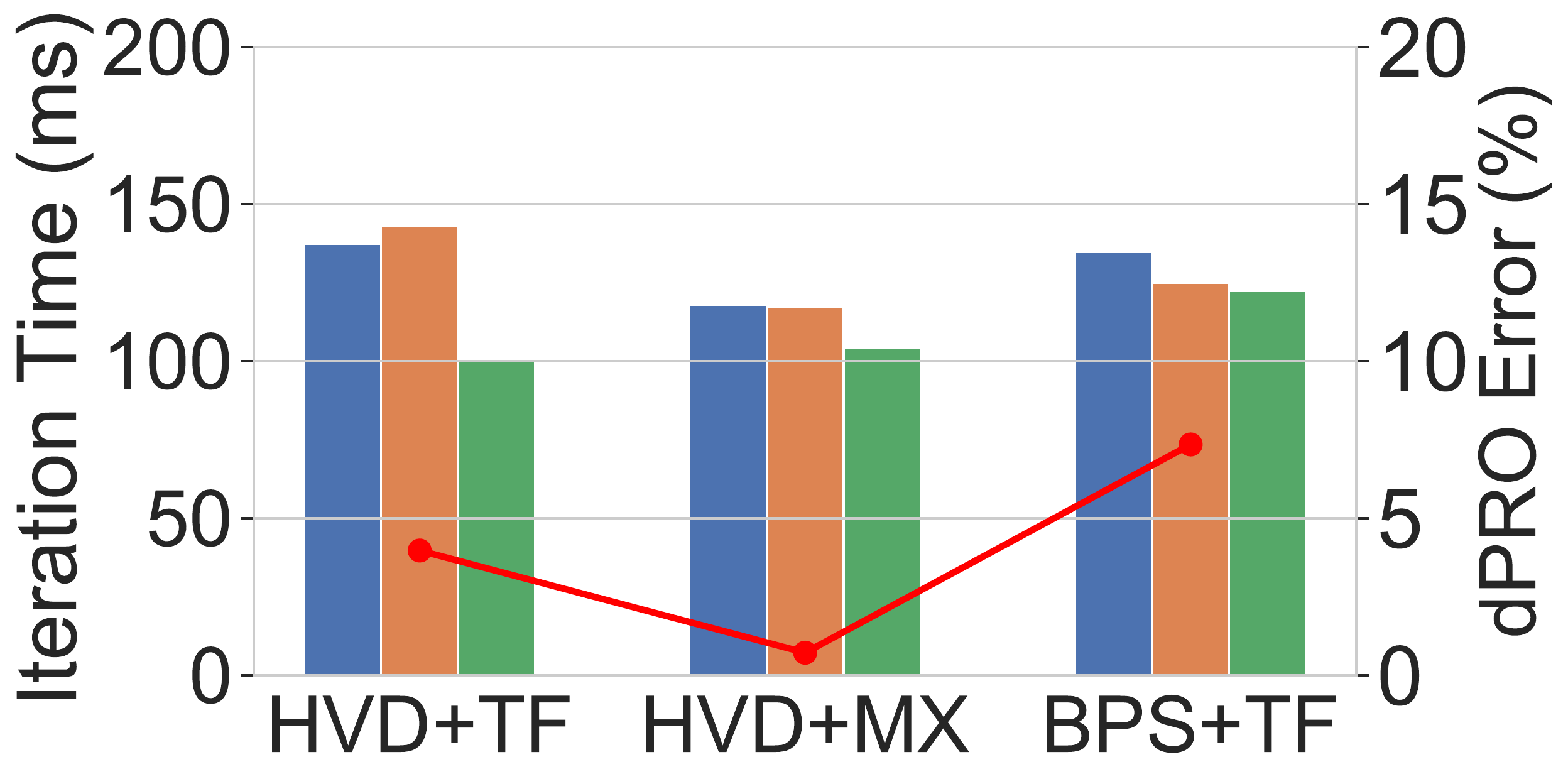}
}
\subfigure[InceptionV3, RDMA]{     
\centering 
\includegraphics[width=0.235\textwidth, trim=0 0 0 0, clip]{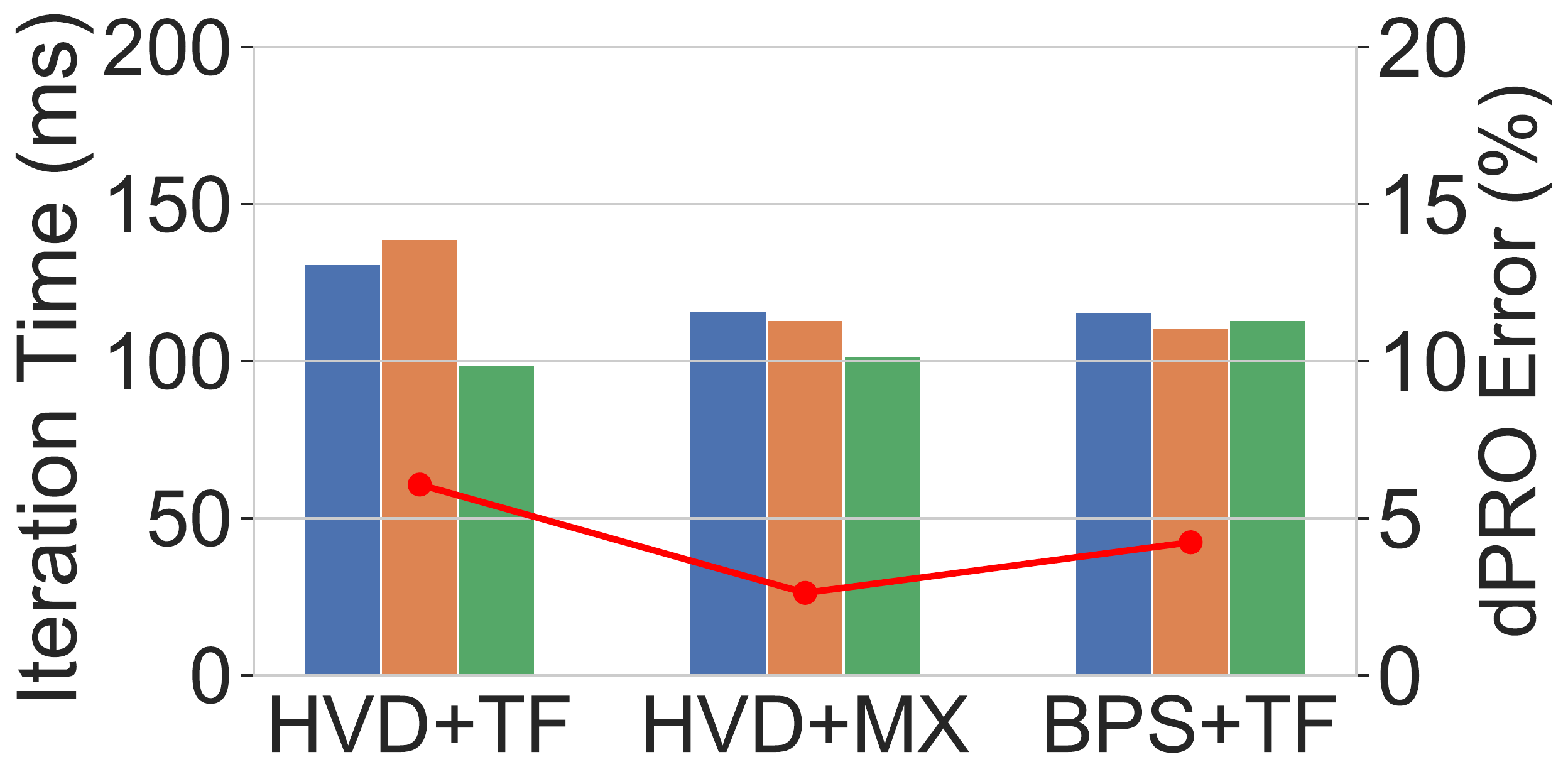}
}
\subfigure[BERT Base, TCP]{     
\centering
\includegraphics[width=0.235\textwidth, trim=0 0 0 0, clip]{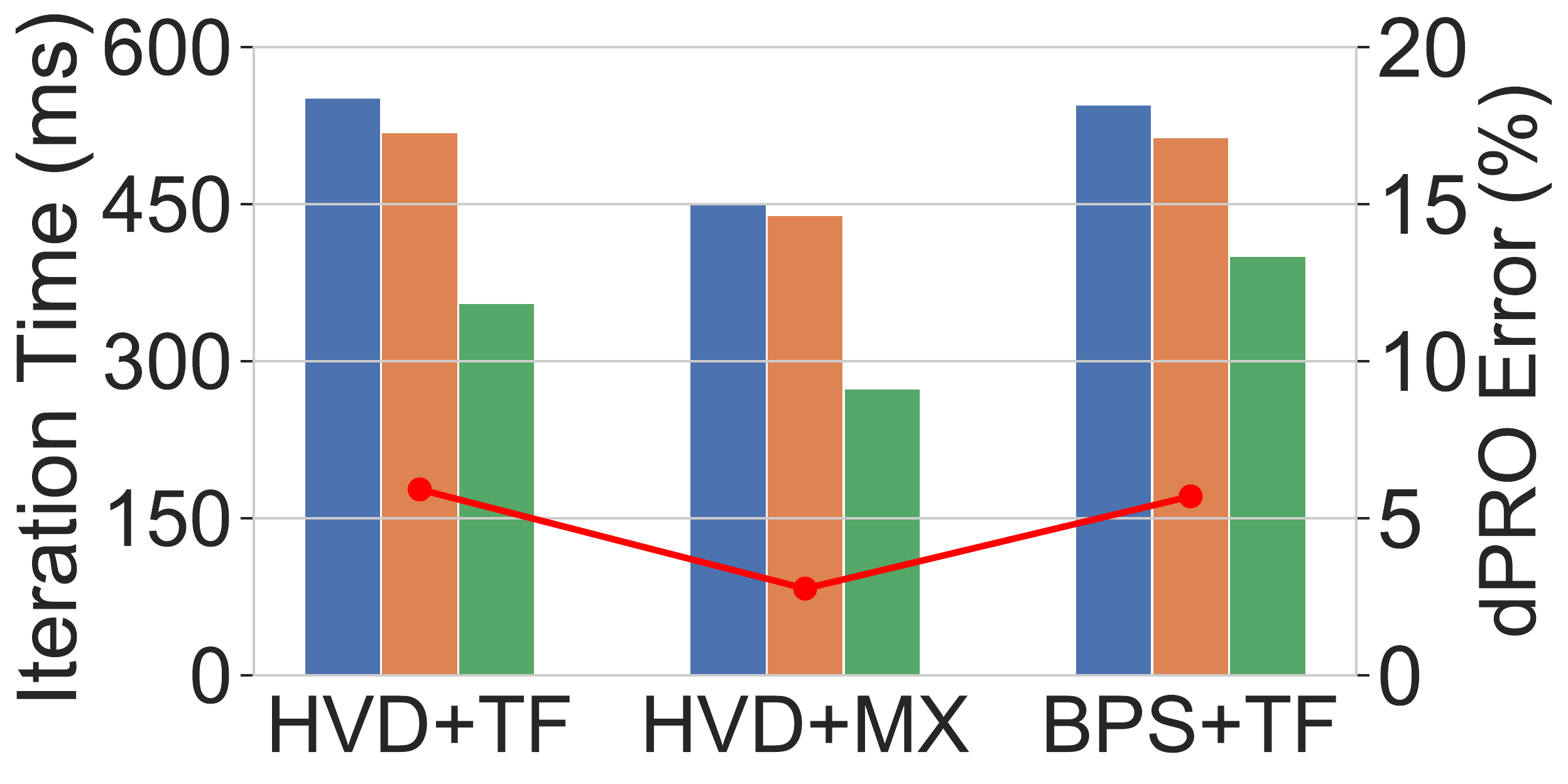}
}
\subfigure[BERT Base, RDMA]{     
\centering
\includegraphics[width=0.235\textwidth, trim=0 0 0 0, clip]{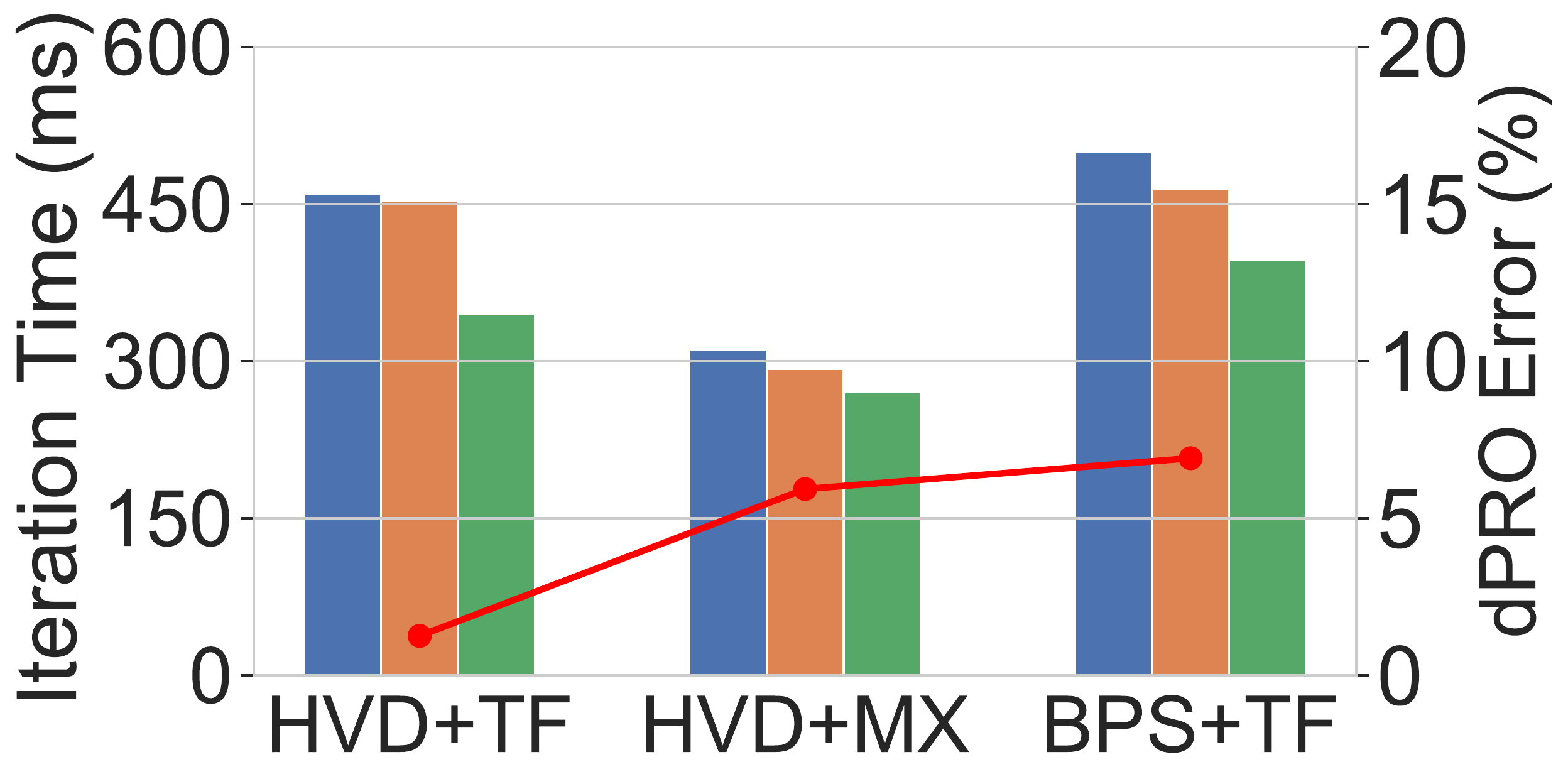}
}
\vspace{-4mm}
\caption{\small{Replay Accuracy: \sysname{} replayer vs.~Daydream simulator 
}}
\label{fig:replay_err}
\end{figure*}

\section{Evaluation}
\subsection{Experiment Setup}
\label{subsec:experiment_setup}
\textbf{Testbed.} We evaluate \sysname{} in a production ML cluster and use up to 128 Tesla V100 32GB GPUs 
(on 16 servers). 
 The GPU servers are equipped with NVLinks and inter-connected with 100Gbps bandwidth using Mellanox CX-5 single-port NICs. We use CUDA v10.2 \cite{cuda} and cuDNN v7.6.5 \cite{cudnn} in our experiments. 
In our default, we use 16 GPUs (on 2 servers).

\noindent \textbf{Benchmarks.} We train 4 DNN models: BERT Base \cite{devlin2018bert} for natural language processing, and ResNet50 \cite{he2016deep}, VGG16~\cite{simonyan2014very} and InceptionV3~\cite{szegedy2016rethinking} for image classification. Each model is trained using various combinations of ML framework, communication library and inter-server connectivity: Horovod TensorFlow (HVD+TF), BytePS TensorFlow (BPS+TF), Horovod MXNet (HVD+MX), BytePS MXNet (BPS+MX), each with TCP or RDMA inter-connectivity.

\noindent \textbf{Baselines.} We compare \sysname{} with the state of the art in various aspects: 1) {\em Daydream} for replay accuracy: 
Daydream \cite{zhu2020daydream} estimates the iteration time of distributed training using a local DFG and inserts one coarse-grained communication op for each tensor, whose communication time is calculated by $tensor\mbox{ }size / bandwidth$. 
2) {\em XLA's default op fusion}, 
which fuses as many computation ops as possible. 
3) {\em Horovod's default tensor fusion}, 
which fuses tensors in intervals of 5ms with fused tensor size upper-bounded by 64MB, as well as 4) {\em Horovod Autotune}, which automatically tunes the interval and upper bound. 
5) {\em BytePS's default setting}, in which
tensors are 
partitioned with a 
partition size of 4MB. 

By default, the batch size is 32 per GPU. Iteration time is averaged over 10 training steps after the warm-up phase. 
More experimental results can be found in the appendix \cite{hu2022appendix}. 


\vspace{-2mm}
\subsection{Replay Accuracy}
Fig.~\ref{fig:replay_err} compares the predicted iteration time by \sysname{} and Daydream's simulator against the real time measured in actual training. \sysname{}'s replay error is less than $5\%$ in most cases, while Daydream's error rate is up to $70.2 \%$. 



\begin{table}[t]
\centering
\small
\caption{\small{Deep Dive of Simulation Error Comparison for TensorFlow Horovod RDMA}}
\vspace{1mm}
\begin{tabular}{|c|c|c|c|c|}
\hline
\multirow{2}*{Model} & \multirow{2}*{Experiment} & \multicolumn{3}{c|}{Time (ms)} \\
\cline{3-5}
~ & ~ & Iteration  & FW & BW\\
\hline
\multirow{3}*{\makecell[c]{ResNet50}} & Ground truth & 138.64 & 34.78 & 71.34 \\
~ & \sysname{} & 142.31 & 34.20 & 70.32\\
~ & Daydream & 109.19 & 34.69 & 70.04 \\
\hline
\multirow{3}*{\makecell[c]{BERT Base}} & Ground truth & 459.62 & 107.49 & 185.66 \\
~ & \sysname{} & 453.83 & 106.58 & 187.05 \\
~ & Daydream & 345.68 & 106.80 & 185.79\\
\hline
\end{tabular}
\label{table:replay_acc_deep_dive}
\end{table}

Table \ref{table:replay_acc_deep_dive} shows 
detailed estimated durations of \textit{forward} 
and \textit{backward} propagation. We observe that both \sysname{} and Daydream predict \textit{forward} and \textit{backward} execution time accurately, so the error mainly arises from the estimation of communication time. Daydream's simple approach for communication time prediction
does not capture effects of message queuing, network and communication protocols, resulting in larger errors in training simulation.

We also note that the overhead introduced by \sysname{}~($5.86\%$ in our experiments) is almost the same as that of ML frameworks' built-in profilers, indicating that our detailed communication profiling introduces little extra overhead.

\vspace{-2mm}
\subsection{Trace Time Alignment}
\begin{figure}[!t]
\subfigure[VGG16, HVD RDMA]{     
\centering
\includegraphics[width=0.225\textwidth, trim=0 0 0 0, clip]{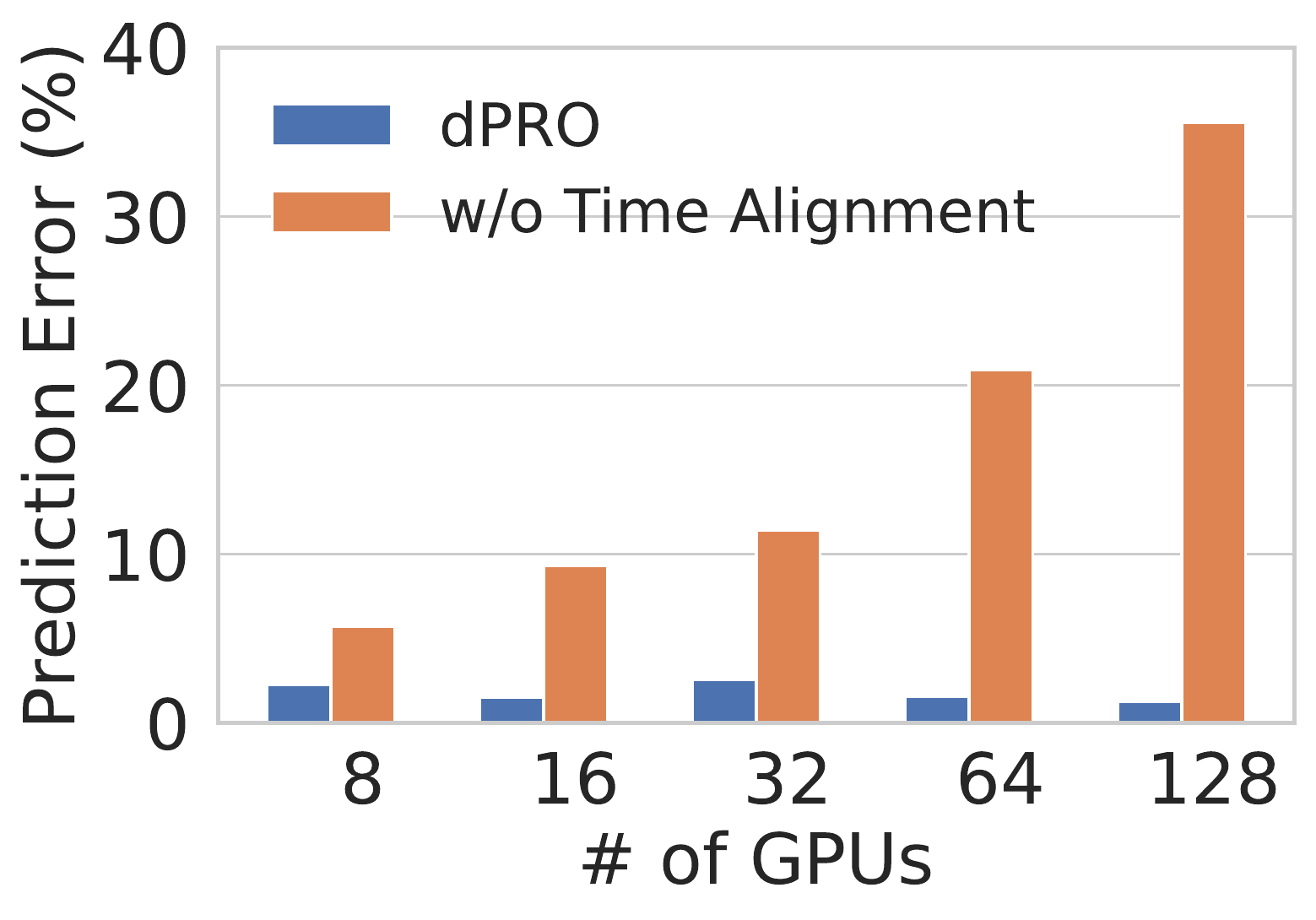}
\label{fig:clock_sync.b}
}
\subfigure[BERT Base, HVD RDMA]{     
\centering
\includegraphics[width=0.225\textwidth, trim=0 0 0 0, clip]{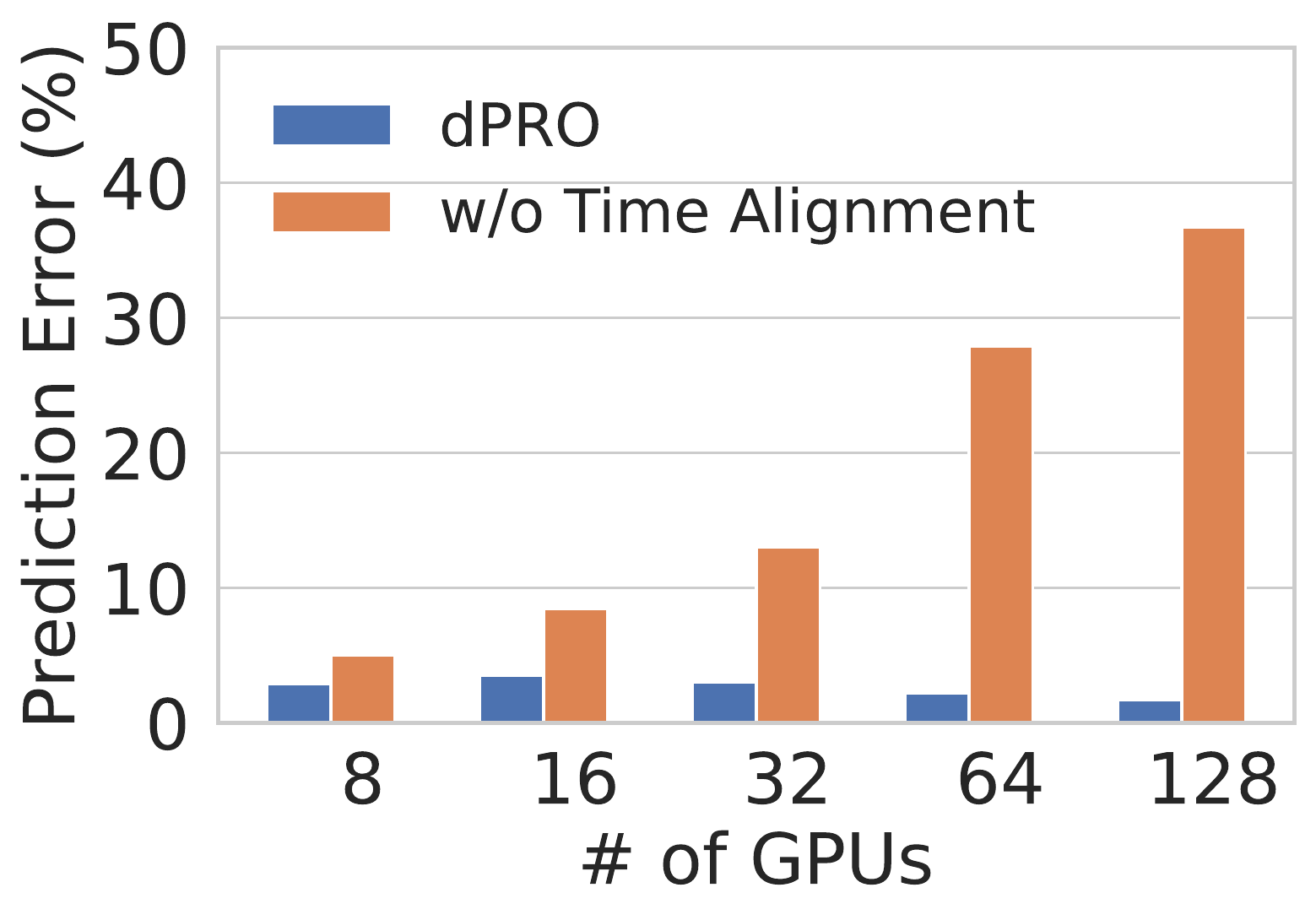}
\label{fig:clock_sync.c}
}
\vspace{-2mm}
\caption{\small{Effects of trace timeline alignment. Workers in the 8-GPU jobs are located on the same physical machine.
}}
\label{fig:clock_sync_scale}
\end{figure}

To evaluate the effect of our 
trace time alignment, we collect traces by running distributed training jobs on different clusters, where NTP is enabled.
Fig.~\ref{fig:clock_sync_scale} shows errors of iteration time estimated by our replayer (as compared to ground truth) 
with and without time alignment. Although workers have been synchronized with NTP or with no clock drift (as in the 8-GPU jobs), inaccuracy of communication traces (\eg, due to RECV op) still leads to large replay errors (up to $36.7\%$), while errors are reduced to $<5\%$ with time alignment. 
Since larger clusters are more likely to experience large clock drifts and queuing delay, the simulation error gap between \sysname{} w/ and w/o trace time alignment becomes significantly larger as the cluster size grows.



\vspace{-2mm}
\subsection{Optimizer Performance}
\begin{figure}[!t]
\centering
\includegraphics[width=0.495\textwidth, trim=0 0 0 0, clip]{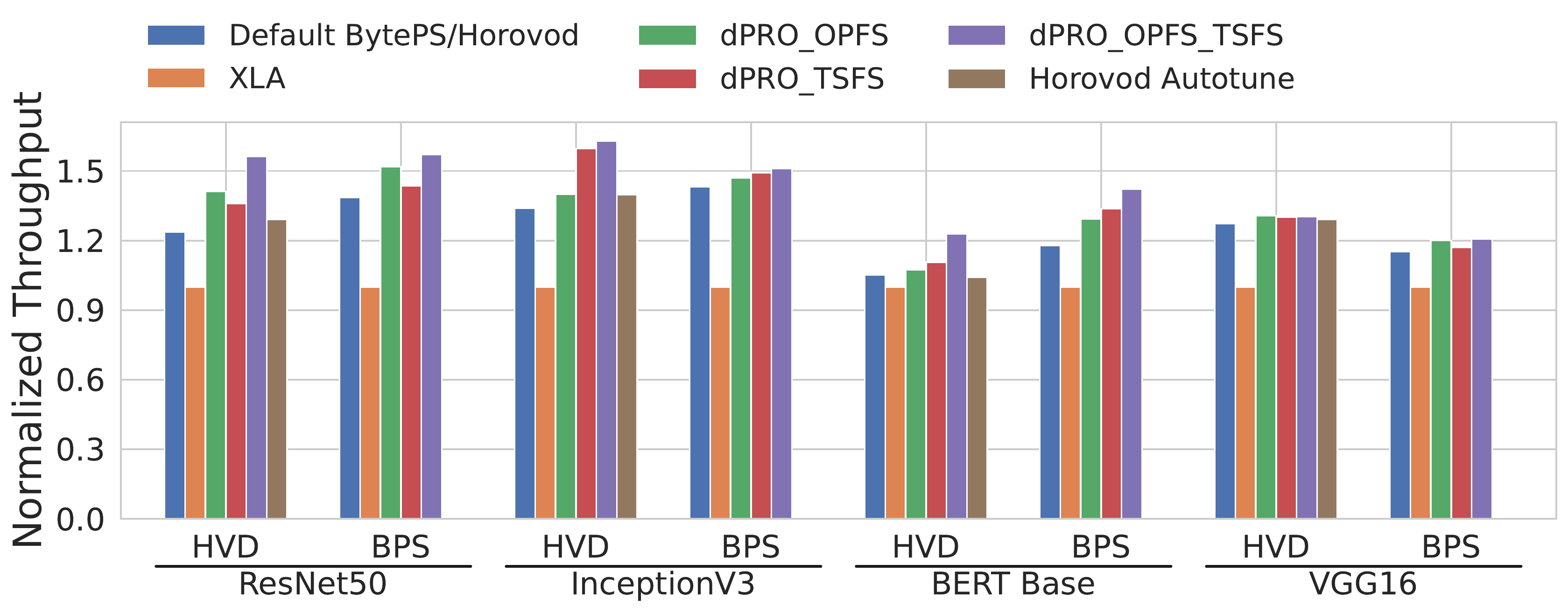}
\label{fig:tsfs_opfs.a}
\vspace{-2mm}
\caption{\small{Performance of op fusion and tensor fusion: training different DNNs on TensorFlow
}}
\label{fig:tsfs_opfs}
\end{figure}

\mypara{Computation op fusion.} 
We first evaluate the performance of op fusion strategies found by the optimizer, temporarily excluding other optimization techniques from the search space. Fig.~\ref{fig:tsfs_opfs} shows 
the actual training throughput when applying the respective strategies in real distributed training. 
\sysname{}'s op fusion  ({\small $\sysname{}\_OPFS$}) yields up to $51.843\%$ speed-up as compared to XLA's default strategies. Although XLA is widely used to accelerate training on a single machine, the results show that in distributed training, simply fusing many ops as with XLA 
may even 
slow down the training progress, due mainly to reduced overlap between computation and communication.

\mypara{Tensor Fusion and Tensor Partition.} We next evaluate the effect of tensor fusion strategies found by \sysname{}'s optimizer ({\small $\sysname{}\_TSFS$}). 
Note that both tensor fusion and tensor partition (BytePS's default partition strategy) are enabled when we use BytePS as the communication library.
As Fig.~\ref{fig:tsfs_opfs} shows,
our strategies achieve up to $19.1\%$ speed-up compared with default Horovod and BytePS.

\mypara{Interaction between op fusion and tensor fusion.} We further evaluate combined op fusion and tensor fusion/partition strategies identified by our optimizer ({\small $\sysname{}\_OPFS\_TSFS$}). 
In Fig.~\ref{fig:tsfs_opfs},
We observe that our combined strategies perform the best in most cases: up to $62.95\%$ acceleration as compared to XLA and up to $26.44\%$ speed-up compared to default Horovod and BytePS. 

\mypara{Integrating memory optimization.} In this experiment, we evaluate the accuracy of peak memory estimation with our replayer. 
Table~\ref{table:mem-est} shows the actual memory consumption and estimated peak memory with our replayer, when training different DNN models on TensorFlow 
with a batch size of 32 per GPU. 
The relative errors across different models are at most 5.25\%. 

\begin{table}[!t]
    \small
    \caption{\small{Memory estimation accuracy}} 
    \vskip 0.05in
    \label{table:mem-est}
    \centering
    \begin{tabular}{|l|c|c|c|} 
        \hline
        Model & Real (GB) & Est. (GB) & Relative Error (\%) \\
        \hline BERT Base & 9.96 & 10.25 & 2.83 \\
        \hline ResNet50 & 5.41 & 5.71 & 5.25 \\
        \hline InceptionV3 & 3.91 & 4.05 & 3.46 \\
        \hline VGG16 & 5.91 & 5.83 & 1.37 \\
        \hline
    \end{tabular}
    \vskip -0.15in
\end{table}

We further investigate the effects of memory optimization (gradient accumulation and re-computation to reduce memory footprint) performed at the start of our search algorithm. Especially, the algorithm evaluates iteration time and memory incurred by the two memory optimizations and selects the one achieving the shorter time under the memory budget.
 Table~\ref{table:mem-opt} presents the results when training BERT Base using a batch size of 64 per GPU, on 16GB V100 GPUs (an OOM error occurs without memory optimization). Re-computation performs better than gradient accumulation in both time and memory consumption in this setting. Moreover, prediction results with our replayer match well the measured data collected in actual training. 

\begin{table}[!t]
    \small
    \vskip -0.05in
    \caption{\small{Iteration time and memory usage: BERT Base (TensorFlow Horovod RDMA) on 16 GPUs with batch size 64 per GPU.}}
    \vskip 0.05in
    \label{table:mem-opt}
    \begin{tabular}{|l|c|c|c|c|} 
        \hline
        \multirow{2}*{Optimization Method} & \multicolumn{2}{c|}{Time (ms)} & \multicolumn{2}{c|}{Memory (GB)} \\ 
        \cline{2-5}
        ~ & Real & Est. & Real & Est. \\ 
        \hline
        w/o optimization & 622.12 & 613.87 & 16.42 & 16.97  \\
        \hline
         Re-computation & 696.13 & 672.60 & 7.43 & 7.20  \\
        \hline
         Gradient Accumulation & 714.22 & 708.57 & 9.96 & 10.25 \\
        \hline
    \end{tabular}
    \vskip -0.15in
\end{table}



\begin{table}[!t]
    \small
    \caption{\small{Time to search optimal op fusion and tensor fusion/partition strategies on TensorFlow BytePS (in hours).}}
    \vskip 0.05in
    \label{table:speedup}
    \centering
    \begin{tabular}{|l|c|c|c|c|} 
        \hline Model & \makecell[c]{Straw-\\man} & \makecell[c]{+Coarsened\\ View} & \makecell[c]{+Partial\\ Replay} & \makecell[c]{+Sym-\\metry} \\
        \hline ResNet50 & 14.60 & 5.35 & 0.91 & 0.29  \\
        \hline VGG16 & 11.97 & 3.74 & 0.71 & 0.04 \\
        \hline InceptionV3 & 16.75 & 6.13 & 1.04 &  0.47 \\
        \hline BERT Base & $>$24 & 22.01 & 3.25 & 0.49  \\
        \hline
    \end{tabular}
    \vskip -0.2in
\end{table}

\mypara{Search speedup.}
We evaluate the effects of our techniques used to accelerate the strategy search process. Table~\ref{table:speedup} gives the algorithm running time, where the search ends when the per-iteration training time evaluated with the identified strategies changes little over 5 search iterations. 
The strawman case is Alg.~\ref{algo_search} without any search speed-up methods, which costs tens of hours. The last three columns show the search time when the respective speed-up method is in place (in addition to the previous one(s)). With more speed-up methods applied, the strategy search time decreases significantly and we can finish the search within a short time. Note that all experimental results we presented earlier are based on strategies found with the speeded-up search.

\begin{figure}[!t]
\subfigure[Replay Accuracy Comparison]{     
\centering
\includegraphics[width=0.49\textwidth, trim=0 0 0 0, clip]{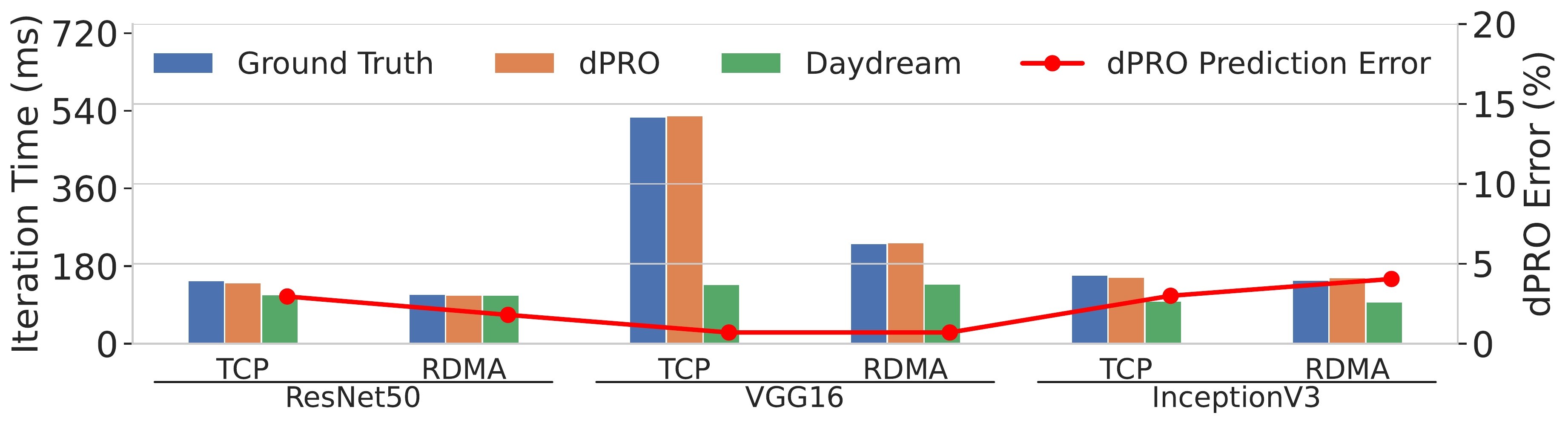}
\label{fig:replay128g}
}
\vspace{-1mm}    
\centering
\includegraphics[width=0.6\textwidth, trim=800 50 0 35, clip]{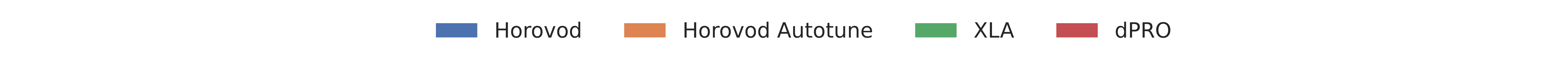}
\subfigure[VGG16]{     
\centering
\includegraphics[width=0.225\textwidth, trim=0 0 0 0, clip]{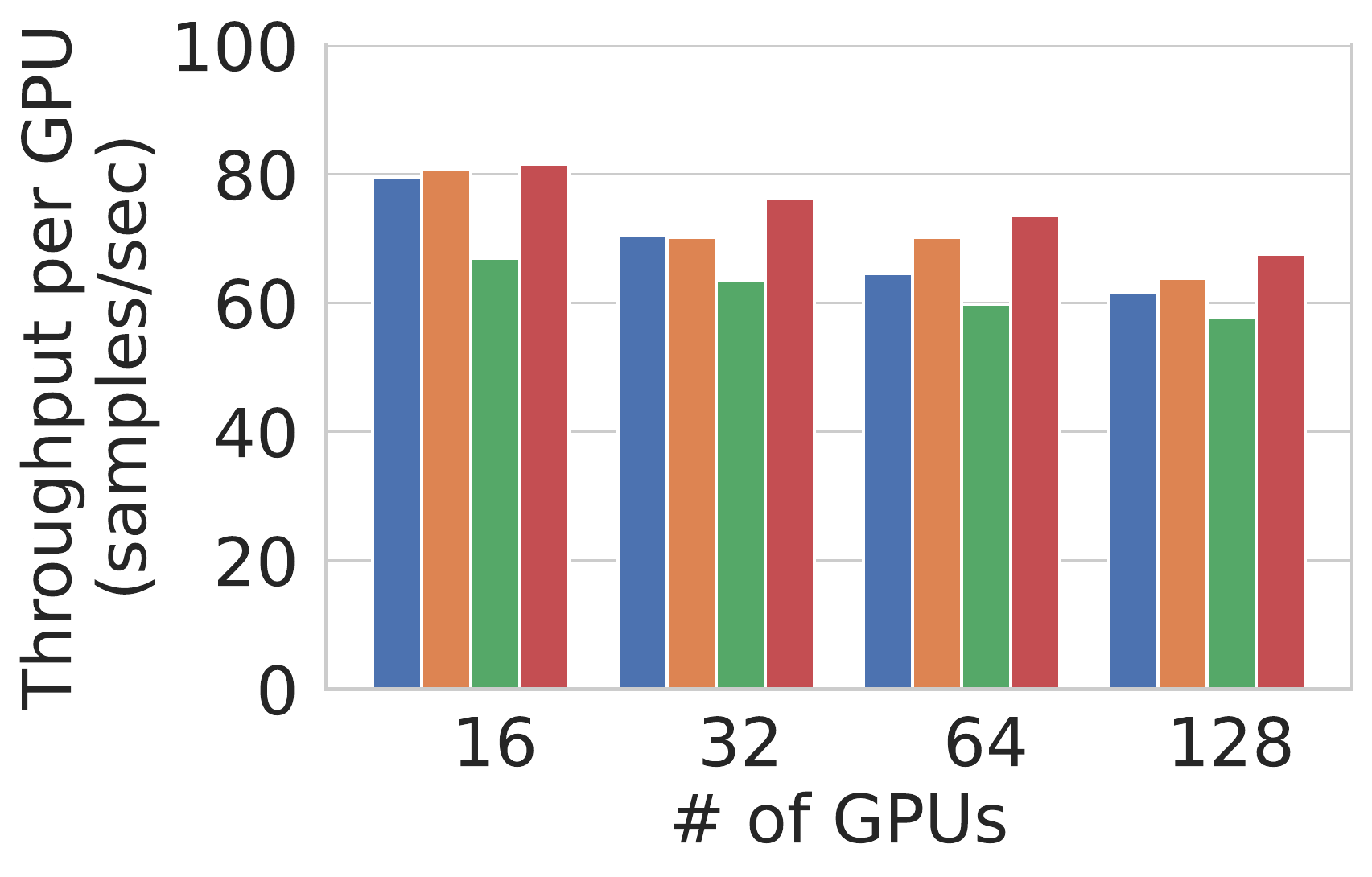}
\label{fig:scale.a}
}
\subfigure[BERT Base]{     
\centering
\includegraphics[width=0.225\textwidth, trim=0 0 0 0, clip]{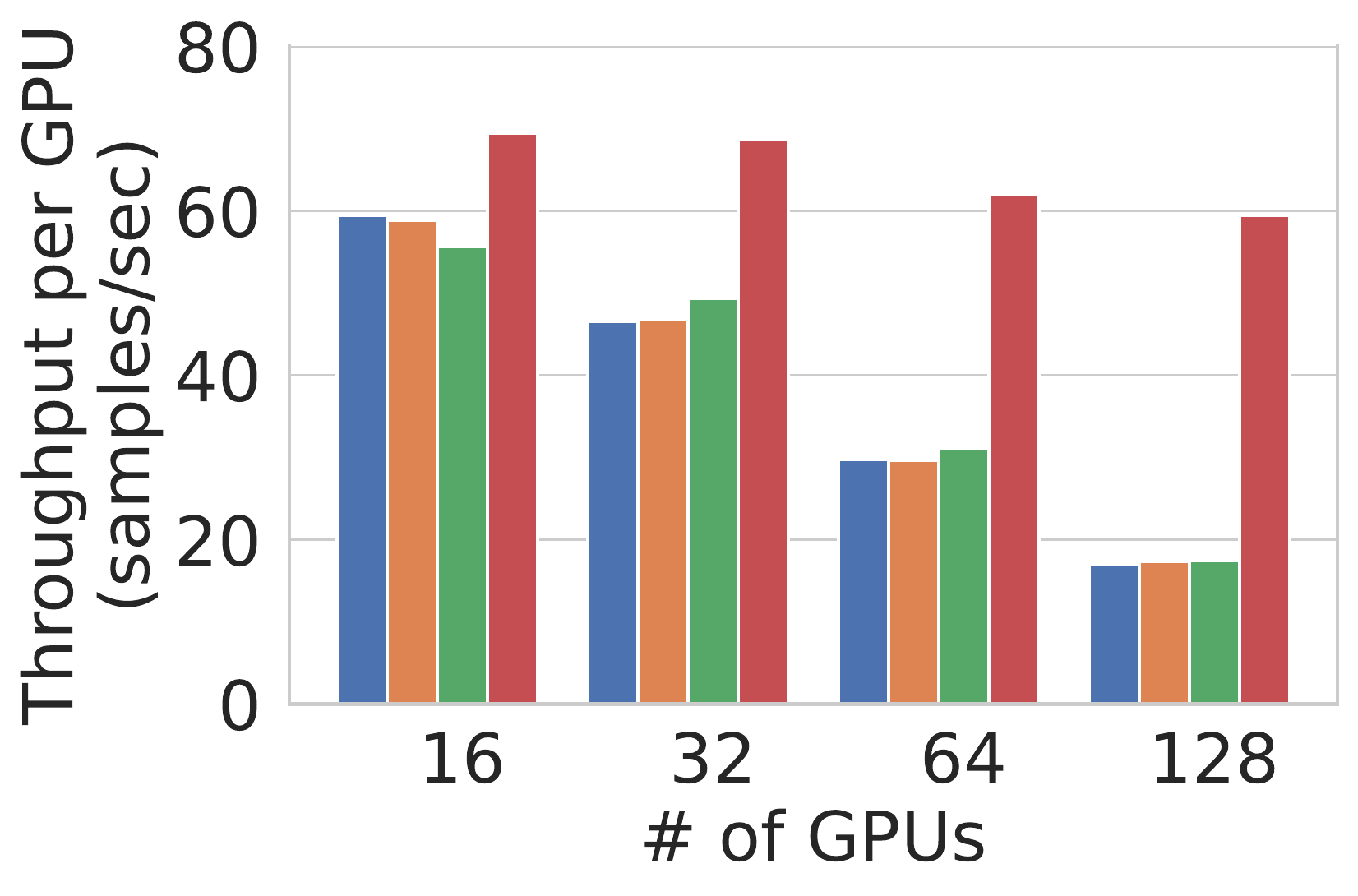}
\label{fig:scale.b}
}
\vspace{-2mm}
\caption{\small{Performance when training DNN models using up to 128 V100 GPUs on TensorFlow + Horovod}}
\label{fig:large_scale}
\end{figure}


\vspace{-2mm}
\subsection{Large-Scale Distributed Training}
We further evaluate \sysname{} on replaying and optimizing distributed DNN training using large-scale clusters.
In Fig.~\ref{fig:large_scale}, we observe the following: (1) as the cluster size grows, DNN training becomes slower since the communication overhead is larger for synchronizing among more GPUs. 
(2) 
Our replayer can still simulate such distributed training accurately, with an error lower than $5\%$ in most cases (up to $5.6 \%$); Daydream's prediction error increases substantially (up to $73.8 \%$) as the cluster size grows. 
(3) \sysname{}'s combined strategies reach the best scalability and yield up to $3.48 \times$ speed-up compared to XLA's default strategies. 

\section{Conclusion}

\sysname{}~is a diagnosis and optimizing toolkit for distributed DNN training. It can simulate distributed training with low error under different configurations and efficiently explore collaborative effects among multiple DNN optimizations. 
The key design points include: 1) building an accurate global DFG by collecting fine-grained traces and aligning timestamps in different devices; 
2) designing an optimization framework to search for combined optimization strategies efficiently. 


\sysname{}, along with the time alignment method, can also be applied to model or pipeline-parallel training, where each local DFG only contains part of ops in the DNN model and the communication topology models transmission of activations between workers. 
We also provide an interface for developers to register custom optimization strategies (e.g., mixed precision), and the optimizer can automatically explore all registered optimizations (see the appendix \cite{hu2022appendix} for more details).


\section{Acknowledgement }

This work was supported by Hong Kong Innovation and Technology Commission’s Innovation and Technology Fund (Partnership Research Programme with ByteDance Limited, Award No. PRP/082/20FX), and grants from Hong Kong RGC under the contracts HKU 17204619, 17208920 and 17207621.

\bibliographystyle{mlsys2022}
\bibliography{reference}




\end{document}